\UseRawInputEncoding
\documentclass[reprint,nofootinbib,showpacs]{revtex4}
\usepackage{graphicx}  
\usepackage{dcolumn}   
\usepackage{bm}        
\usepackage{amssymb}
\usepackage{multirow}
\usepackage{amsmath}
\usepackage{xcolor}
\usepackage{hyperref}
\usepackage[figurename=Figure]{caption}
\usepackage{caption}
\usepackage{subcaption}

\begin{document}

\title{Phenomenological study of two minor zeros in neutrino mass matrix using trimaximal mixing}
 \author{Iffat Ara Mazumder${}$}
 \email{iffat\_rs@phy.nits.ac.in}
 \author{Rupak~Dutta${}$}
 \email{rupak@phy.nits.ac.in}
 \affiliation{${}$ National Institute of Technology Silchar, Silchar 788010, India}

\begin{abstract}
We study the phenomenological implications of two minor zeros in neutrino mass matrix using trimaximal mixing matrix. In this context, we
analyse fifteen possible cases of two minor zeros in neutrino mass matrix and found only two cases, namely Class $A_1$ and Class $A_2$, that
are compatible with the present neutrino oscillation data. We present correlations of several neutrino oscillation parameters and give
prediction of
the total neutrino mass, the values of effective Majorana mass, the effective electron anti-neutrino mass and CP violating Majorana phases for
these two classes. We also explore the degree of fine tuning in the elements of neutrino mass matrix for the allowed classes. Moreover, we 
propose a flavor model within the seesaw model along with $Z_{8}$ symmetry group that can generate such classes.
\end{abstract}

\maketitle
\section{Introduction}
Although no new particles beyond the Standard Model~(SM) have been discovered in any experiments so far, but the neutrino oscillation
phenomena observed in several experiments~\cite{Super-Kamiokande:1998kpq, Fogli:2001vr} indirectly confirms the existence of a more global
theory beyond the SM. The non zero neutrino mass as confirmed by neutrino oscillation experiments can not be explained by the SM and its 
origin is one of the most fundamental questions in particle physics today. In the flavor basis where the charged lepton mass matrix is 
diagonal, the neutrino mass matrix $M$ can be parametrized in terms of nine free parameters, namely three masses~$(m_1,\,m_2,\,m_3)$,
three mixing angles~$(\theta_{12},\,\theta_{13},\,\theta_{23})$, two Majorana CP violating phases~$(\alpha,\,\beta)$ and a Dirac CP
violating phase~$(\delta)$. Although the mixing angles and the mass squared differences are known to a very good accuracy from currently
available experimental data, there are still many open questions in the neutrino sector. For instance, the absolute mass scale of three
neutrinos, the sign of $\Delta m_{32}^2$, whether neutrinos are Dirac or Majorana particle, the three CP violating phases are still unknown. 
With the experimental values of the mass squared differences and the mixing angles as the input parameters, we still have four unknown 
parameters in the neutrino mass matrix $M$. There exists several schemes in literature by which one can reduce the number of free
parameters in the neutrino mass matrix. The general idea is to assume that some matrix elements will be dependent on other matrix elements. 
For instance, one can assume certain matrix element to be zero or certain combination of elements of the neutrino mass matrix to be zero 
that can be caused by some underlying flavor symmetry.

The exact mechanism of the origin of neutrino masses is still unknown. Out of many proposed theoretical models, the seesaw mechanism of 
either type-I or type-II looks more promising. The seesaw mechanism not only helps in understanding the scale of neutrino mass but also
provide the necessary ingredients to realize texture zero patterns in the neutrino mass matrix.
Within the framework of Type-I seesaw mechanism~\cite{Minkowski:1977sc,Mohapatra:1979ia}, the neutrino mass matrix is written as 
\begin{eqnarray}
M=-M_DM_R^{-1}M_D^{T}\,,
\end{eqnarray}
where $M_D$ is the $3\times 3$ Dirac mass matrix that links $(\nu_e,\,\nu_{\mu},\,\nu_{\tau})$ to their right handed singlet counterpart
and $M_R$ is the $3\times 3$ Majorana mass matrix of the right handed neutrinos. It is worth mentioning
that by considering the suitable position of zeros in $M_D$ and $M_R$, one can, in principle, get the desired zero textures of the neutrino 
mass matrix. It should also be noted that zeros in $M_D$ and $M_R$ not only give rise to zeros in the mass matrix but also give rise to
zero minors in the neutrino mass matrix. It was also pointed out in Ref.~\cite{Lashin:2009yd,Mazumder:2022ywa} that if $M_D$ is diagonal, then zeros in $M_R$ can give rise 
to minor zero pattern in the neutrino mass matrix $M$.

There exists numerous literatures where the form and texture of the neutrino mass matrix have been explored. It was found that texture with 
more than two zeros in the neutrino mass matrix can not accommodate the latest neutrino oscillation data.
The phenomenology of two texture zeros~\cite{Frampton:2002yf,Kumar:2017hjn,Xing:2002ap,Lavoura:2004tu,Dev:2006qe,Kumar:2011vf,Fritzsch:2011qv,
Ludl:2011vv,Meloni:2012sx,Grimus:2012zm,Gautam:2016qyw,Channey:2018cfj,Singh:2019baq} have been explored at large scale in the literature with
Pontecorvo~Maki~Nakagawa~Sakata~(PMNS) and trimaximal~(TM) mixing matrix. 
It should be noted that out of fifteen possible cases of two texture zeros only seven patterns are favored by the recent experimental 
data.~\cite{Frampton:2002yf} in case of PMNS mixing matrix. With TM mixing matrix, however, only two patterns are found to be compatible
with the oscillation data.~\cite{Kumar:2017hjn,Gautam:2016qyw}. 
In Ref.~\cite{Lavoura:2004tu}, the authors investigated viable textures with two zeros in the inverted neutrino mass matrix and seven
such patterns were shown to be allowed. However, these textures do not apply in the case of non-invertible neutrino mass matrix $M$. 
Also, in Ref.~\cite{Lashin:2011dn,Gautam:2018izb}, authors have studied the phenomenological implications of one texture zero in neutrino 
mass matrix and found all six possible patterns to be compatible with the data. Moreover, in 
Ref.~\cite{Lashin:2007dm,Lashin:2009yd,Dev:2010if,Dev:2010pf,Araki:2012ip,Mazumder:2022ywa} and Ref.~\cite{Liao:2013saa,Dev:2013xca,
Wang:2013woa,Whisnant:2015ovx,Dev:2015lya,Wang:2016tkm} the authors have explored the phenomenology of vanishing minor and cofactor zero in 
the neutrino mass matrix. The class of one and two independent zero minors textures in the neutrino mass matrix were explored 
in Refs.~\cite{Lashin:2009yd,Lashin:2007dm} with PMNS mixing matrix. It was shown that out of fifteen possible two minor zero patterns, only 
seven patterns are 
viable and all the six one minor zero patterns are compatible with the data. Similarly, for TM mixing matrix along with one minor 
zero condition~\cite{Mazumder:2022ywa}, all the six possible patterns are found to be allowed. In Ref.~\cite{Dev:2010pf}, the authors have 
used the tribimaximal~(TB) mixing matrix and showed that five classes of texture zeros or vanishing minor can accommodate the neutrino 
oscillation data. 

Although TB mixing pattern was studied extensively in literature, it, however, was ruled out due to a non zero value of the reactor mixing 
angle~$\theta_{13}$ confirmed by several experiments such as
T2K~\cite{T2K:2011ypd}, MINOS~\cite{MINOS:2011amj}, Double Chooz~\cite{DoubleChooz:2011ymz} and Daya Bay~\cite{DayaBay:2012fng}.
In order to accommodate a non zero value of $\theta_{13}$, TM mixing matrix was constructed by multiplying the TB mixing matrix by an unitary 
matrix~\cite{Kumar:2010qz,He:2011gb,Grimus:2008tt}. If the first column of the TM mixing matrix is 
identical to the first column of TB mixing matrix, it is called TM$_1$ mixing matrix and if the second column is identical to that of the 
TB mixing matrix it is called TM$_2$ mixing matrix.
In this paper we analyze the compatibility of two minor zero neutrino Majorana textures with the recent experimental data using  
TM$_1$ and TM$_2$ mixing matrix.
By using two minor zero conditions one can reduce the number of free parameters in the model. We analyse fifteen possible cases of two minor 
zeros in neutrino mass matrix and found only two cases, namely class $A_1$ and class 
$A_2$, that are compatible with the present neutrino oscillation data. In order to study the detail feature of the mass matrix elements we 
first perform a $\chi^2$ analysis of the allowed two minor zero textures. We use five observables in our $\chi^2$ analysis, namely three
mixing angles and the two mass squared differences. We do not use Dirac CP violating phase $\delta$ in our $\chi^2$ analysis as its value
has not been measured yet. We also find the degree of fine tuning in the neutrino mass matrix elements for each 
allowed textures. Moreover, We give our model predictions of the unknown Dirac and Majorana 
CP violating phases, the total neutrino mass, effective Majorana mass and the effective electron anti-neutrino mass for these two classes.

We organize the paper as follows. We start with TM$_1$ and TM$_2$ mixing matrix and find the elements of the neutrino mass matrix in 
Section.~\ref{section:2}. We then write down $\theta_{13}$, $\theta_{23}$, $\theta_{12}$, Dirac CP violating phase $\delta$, effective
electron anti-neutrino mass $m_{\nu}$ and the effective Majorana mass term $M_{ee}$ in terms of the unknown parameters $\theta$ and $\phi$
of the TM mixing matrix. In
Section.~\ref{section:3}, we describe the formalism of two minor zeros in neutrino mass matrix and identify all the possible cases
of two minor zero in neutrino mass matrix. In Section.~\ref{section:4}, we discuss the phenomenology of the allowed two minor zero classes and obtain the best fit values of all the oscillation parameters with our $\chi^2$ analysis. Also we report the degree of fine-tuning in the neutrino mass matrix elements. In Section.~\ref{section:5}, we present the symmetry realization of our model
and conclude in Section.~\ref{section:6}.

\section{Neutrino mass matrix}
\label{section:2}
TM$_1$ and TM$_2$  mixing matrix, constructed by multiplying the TB mixing matrix by an unitary matrix, can be written as
\begin{equation} 
\label{eq:2}
  U_{TM_1}=\begin{pmatrix}
 \sqrt{\frac{2}{3}} & \frac{1}{\sqrt{3}}\cos\theta & \frac{1}{\sqrt{3}}\sin\theta\\
 -\frac{1}{\sqrt{6}}& \frac{\cos\theta}{\sqrt{3}}-\frac{e^{i\phi}\sin\theta}{\sqrt{2}}&\frac{\sin\theta}{\sqrt{3}}+\frac{e^{i\phi}\cos\theta}{\sqrt{2}}\\
 -\frac{1}{\sqrt{6}}& \frac{\cos\theta}{\sqrt{3}}+\frac{e^{i\phi}\sin\theta}{\sqrt{2}}&\frac{\sin\theta}{\sqrt{3}}-\frac{e^{i\phi}\cos\theta}{\sqrt{2}}\,
\end{pmatrix}. 
 \end{equation}
and
\begin{equation} 
\label{eq:3}
  U_{TM_2}=\begin{pmatrix}
 \sqrt{\frac{2}{3}}\cos\theta & \frac{1}{\sqrt{3}} & \sqrt{\frac{2}{3}}\sin\theta\\
 -\frac{\cos\theta}{\sqrt{6}}+\frac{e^{-i\phi}\sin\theta}{\sqrt{2}}& \frac{1}{\sqrt{3}}&-\frac{\sin\theta}{\sqrt{6}}-\frac{e^{-i\phi}\cos\theta}{\sqrt{2}}\\
 -\frac{\cos\theta}{\sqrt{6}}-\frac{e^{-i\phi}\sin\theta}{\sqrt{2}}& \frac{1}{\sqrt{3}}&-\frac{\sin\theta}{\sqrt{6}}+\frac{e^{-i\phi}\cos\theta}{\sqrt{2}}\,
\end{pmatrix}. 
 \end{equation}
where $\theta$ and $\phi$ are two free parameters. 
In the flavor basis where the charged lepton mass matrix is diagonal, the symmetric neutrino mass matrix $M$ can be expressed as
\begin{equation} 
\label{eq:4}
 (M)_{\rho\sigma}=(V\,M_{diag}\,V^{T})_{\rho\sigma}\,\, {\rm with}\,\, \rho\,, \sigma = e\,,\mu\,,\tau\,,
\end{equation}
where $M_{diag} = {\rm diag}(m_1, m_2, m_3)$ is the diagonal matrix containing three mass states. The lepton flavor mixing matrix $V$ can be
expressed as $V=U_{TM}\,P$, where the diagonal phase matrix $P$ can be written as
\begin{equation} 
\label{eq:5}
 P=\begin{pmatrix}
  1& 0 & 0\\
 0 & e^{i\alpha} &0\\
 0 & 0 & e^{i\beta}\
\end{pmatrix}\,
\end{equation}
that contains two CP violating Majorana phases $\alpha$ and $\beta$. 
For completeness, we report all the elements of the neutrino mass matrix for the TM$_1$ and TM$_2$ mixing matrix in Eq.~\ref{eq:55} and 
Eq.~\ref{eq:56} of appendix \ref{app}.

We can express the neutrino oscillation parameters such as the three mixing angles $\theta_{12}$, $\theta_{23}$, $\theta_{13}$, the Jarlskog 
invariant $J$~\cite{Jarlskog:1985ht} and the Dirac CP violating phase $\delta$ in terms of the two unknown parameters $\theta$ and $\phi$ of 
the trimaximal mixing matrix. For TM$_1$ mixing matrix, we have

\begin{eqnarray} 
\label{eq:7}
&& s_{12}^2=\frac{|(U_{12})_{TM_1}|^2}{1-|(U_{13})_{TM_1}|^2} = 1-\frac{2}{3-\sin^2\theta}\,, \nonumber \\
&& s_{23}^2=\frac{|(U_{23})_{TM_1}|^2}{1-|(U_{13})_{TM_1}|^2} = \frac{1}{2}\Big(1+\frac{\sqrt{6}\sin2\theta \cos\phi}{3-\sin^2\theta}\Big)\,,
\nonumber \\
&& s_{13}^2=|(U_{13})_{TM_1}|^2 = \frac{1}{3}\sin^2\theta\,, \nonumber \\
&& J=\frac{1}{6\sqrt{6}}\sin2\theta \sin\phi\,, \nonumber \\
&& \csc^2\delta=\csc^2\phi-\frac{6\sin^{2}2\theta \cot^2\phi}{(3-\sin^2\theta)^2}\,
\end{eqnarray}
and for TM$_2$ mixing matrix, we have
\begin{eqnarray}
\label{eq:14}
&& s_{12}^2=\frac{|(U_{12})_{TM_2}|^2}{1-|(U_{13})_{TM_2}|^2} = \frac{1}{3-2\sin^2\theta}\,, 
\nonumber \\
&& s_{23}^2=\frac{|(U_{23})_{TM_2}|^2}{1-|(U_{13})_{TM_2}|^2} = \frac{1}{2}\Big(1+\frac{\sqrt{3}\sin2\theta \cos\phi}{3-2\sin^2\theta}\Big)\,,
\nonumber \\
&& s_{13}^2=|(U_{13})_{TM_2}|^2 = \frac{2}{3}\sin^2\theta\,, \nonumber \\
&& J=\frac{1}{6\sqrt{3}}\sin2\theta \sin\phi\,, \nonumber \\
&& \csc^2\delta=\csc^2\phi-\frac{3\sin^{2}2\theta \cot^2\phi}{(3-2\sin^2\theta)^2}\,,
\end{eqnarray}
where $s_{ij}=\sin\theta_{ij}$ and $c_{ij}=\cos\theta_{ij}$ for $i,j=1,2,3$.

There are several experiments that can, in principle, put constraints on the neutrino mass scale. The $\beta$ decay experiment performed at 
Karlsruhe Tritium Neutrino (KATRIN) experiment can measure the effective electron anti-neutrino mass by studying the endpoint region of the 
$\beta$ decay spectrum. The current upper bound of the effective electron anti-neutrino mass is reported to be $m_{\nu} < 0.8\,{\rm eV}$ at 
$90\%$ confidence level~\cite{KATRIN:2021uub}. In future KATRIN is expected to reach the mass sensitivity up to $0.2\,{\rm eV}$. 
The next generation $\beta$ decay experiment such as Project $8$ experiment~\cite{Project8:2022wqh} is designed to reach the mass
sensitivity up to $0.04\,{\rm eV}$.
Similarly,
the effective Majorana mass term $|M_{ee}|$ can be obtained from the neutrinoless double beta decay experiment. The current upper bound on 
the value of $M_{ee}$ reported by GERDA experiment obtained by using $^{48}$Ca isotope is $M_{ee} < (0.079 - 0.18)\,{\rm eV}$~
\cite{GERDA:2013vls}. The KamLAND-Zen and EXO-200 experiments~\cite{KamLAND-Zen:2012mmx,EXO-200:2019rkq} on $^{136}$Xe isotope reported the 
upper limit on $M_{ee}$ to be $M_{ee} < (0.061 - 0.165)\,{\rm eV}$ and $M_{ee} < (0.09 - 0.29)\,{\rm eV}$, respectively. It is expected
that the next generation experiments can reach the mass sensitivity upto $(5-20)\,{\rm meV}$.
There are also several results related to the total neutrino mass coming from various cosmological observations. The Planck satellite 
reported the upper limit on the total neutrino mass combining BAO data with CMB data to be $\sum\,m_i < 0.12\,{\rm eV}$ at $95\%$ confidence 
level ~\cite{Zhang:2020mox}. The stringent limit on the absolute neutrino mass is obtained by combining CMB lensing and galaxy clustering data
and it is found to be $\sum\,m_i < 0.09\,{\rm eV}$ ~\cite{Palanque-Delabrouille:2019iyz}. 

We can write the effective Majorana mass $|M_{ee}|$ and effective electron anti-neutrino mass $m_{\nu}$ for the TM$_1$ mixing matrix as
\begin{equation} 
\label{eq:11}
|M_{ee}|=\Big|\frac{1}{3}(2m_1+ m_2\cos^2\theta  e^{2i\alpha} + m_3\sin^2\theta  e^{2i\beta})\Big|.
\end{equation}
\begin{equation} 
\label{eq:12}
 m_{\nu}^2=\sum\limits_{i=1}^{3}m^{2}_{i}U_{ie}^2=\frac{1}{3}(2m_1^2+ m_2^2\cos^2\theta + m_3^2\sin^2\theta).
\end{equation}
For TM$_2$ mixing matrix, $|M_{ee}|$ and $m_{\nu}$ can be expressed as 
  \begin{equation}
\label{eq:17}
  |M_{ee}|=\Big|\frac{1}{3}(2m_1\cos^2\theta + m_2 e^{2i\alpha} + 2m_3\sin^2\theta  e^{2i\beta})\Big|.
 \end{equation}
 \begin{equation} 
\label{eq:18}
  m_{\nu}^2=\frac{1}{3}(2m_1^2\cos^2\theta + m_2^2+ 2m_3^2\sin^2\theta).
 \end{equation}
The most stringent constraint on any model required to fit the data comes from the ratio of the absolute values of the solar and atmospheric 
mass squared differences characterized by
\begin{eqnarray}
r\equiv \Big|\frac{\Delta m_{21}^2}{\Delta m_{32}^2}\Big|\,,
 \end{eqnarray}
where $\Delta m_{21}^2$ and $\Delta m_{32}^2$ represent solar and atmospheric mass squared differences, respectively. With the latest
experimental values of $\Delta m_{21}^2$ and $\Delta m_{32}^2$, one can estimate the value of $r$ to be $(2.95\pm0.08)\times10^{-2}$.

\section{Two minor zeros in neutrino mass matrix}\label{section:3}
The neutrino mass matrix constructed using the trimaximal mixing matrix is a $3 \times 3$ symmetric matrix and has six independent entries. 
Hence we have six independent minors corresponding to each independent entries in the mass matrix. There are total ${^6}C_2$ or $15$ possible
ways to have two minor zeros in the mass matrix. All the fifteen possible patterns of two minor zero in neutrino mass matrix are listed in 
Table.~\ref{tab:1}. We denote the minor corresponding to $ij^{th}$ element of $M_{ij}$ as $C_{ij}$. 
\begin{table}[h!]
\begin{tabular}{|c|c|}
 \hline
Class &Constraining equations\\
\hline
$A_1$&$C_{33}=0$,$ C_{32}=0$\\
\hline
$A_2$&$C_{22}=0$,$ C_{32}=0$\\
\hline
$B_3$&$C_{33}=0$,$ C_{31}=0$\\
\hline
$B_4$&$C_{22}=0$,$ C_{21}=0$\\
\hline
$B_5$&$C_{33}=0$,$ C_{12}=0$\\
\hline
$B_6$&$C_{22}=0$,$ C_{13}=0$\\
\hline
D&$C_{33}=0$,$ C_{22}=0$\\
\hline
$S_1$&$C_{31}=0$,$ C_{11}=0$\\
\hline
$S_2$&$C_{21}=0$,$ C_{11}=0$\\
\hline
$S_3$&$C_{13}=0$,$ C_{12}=0$\\
\hline
$F_1$&$C_{33}=0$,$ C_{11}=0$\\
\hline
$F_2$&$C_{22}=0$,$ C_{11}=0$\\
\hline
$F_3$&$C_{32}=0$,$ C_{11}=0$\\
\hline
$F_4$&$C_{31}=0$,$ C_{32}=0$\\
\hline
$F_5$&$C_{21}=0$,$ C_{32}=0$\\
\hline
\hline
\end{tabular}
\caption{Two minor zero patterns. }
\label{tab:1}
\end{table}

In terms of neutrino mass matrix elements, the conditions for two minor zero can be written as
\begin{eqnarray} 
\label{eq:19}
M_{a\,b}M_{c\,d}-M_{u\,v}M_{w\,x}=0\,, \nonumber\\
M_{a^{'}\,b^{'}}M_{c^{'}\,d^{'}}-M_{u^{'}\,v^{'}}M_{w^{'}\,x^{'}}=0\,.
\end{eqnarray}
We can write Eq.~\ref{eq:19} in terms of a complex equation as 
\begin{eqnarray} 
\label{eq:20}
 && m_{1}m_{2}X_{3}e^{2i\alpha}+m_{2}m_{3}X_{1}e^{2i(\alpha+\beta)}+m_{3}m_{1}X_{2}e^{2i\beta}=0\,,\nonumber\\
 && m_{1}m_{2}Y_{3}e^{2i\alpha}+m_{2}m_{3}Y_{1}e^{2i(\alpha+\beta)}+m_{3}m_{1}Y_{2}e^{2i\beta}=0\,,
\end{eqnarray}
where 
\begin{eqnarray}
 \label{eq:21}
 && X_{k}=(U_{ai}U_{bi}U_{cj}U_{dj}-U_{ui}U_{vi}U_{wj}U_{xj})+(i\leftrightarrow j)\,, \nonumber\\
 && Y_{k}=(U_{a^{'}i}U_{b^{'}i}U_{c^{'}j}U_{d^{'}j}-U_{u^{'}i}U_{v^{'}i}U_{w^{'}j}U_{x^{'}j})+(i\leftrightarrow j)\,,
\end{eqnarray}
with $(i,j, k)$ as the cyclic permutation of $(1,2,3)$.
Using Eq.~\ref{eq:20}, one can write the mass ratios as
\begin{eqnarray} 
 \label{eq:22}
&& \frac{m_1}{m_2}e^{-2i\alpha}=\frac{X_{3}Y_{1}-X_{1}Y_{3}}
{X_{2}Y_{3}-X_{3}Y_{2}}\,, \nonumber \\
&& \frac{m_1}{m_3}e^{-2i\beta}=\frac{X_{1}Y_{2}-X_{2}Y_{1}}{X_{2}Y_{3}-X_{3}Y_{2}}\,, \nonumber \\
&& \frac{m_3}{m_2}e^{-2i(\alpha-\beta)}=\frac{X_{3}Y_{1}-X_{1}Y_{3}}{X_{1}Y_{2}-X_{2}Y_{1}}\,.
\end{eqnarray}
Similarly, the CP-violating Majorana phases can be written as 
\begin{eqnarray} 
 \label{eq:23}
 &&\alpha=-\frac{1}{2}\arg(\frac{X_{3}Y_{1}-X_{1}Y_{3}}{X_{2}Y_{3}-X_{3}Y_{2}})\,, \nonumber \\
 &&\beta=-\frac{1}{2}\arg(\frac{X_{1}Y_{2}-X_{2}Y_{1}}{X_{2}Y_{3}-X_{3}Y_{2}})\,.
\end{eqnarray}

The value of $m_1$, $m_2$ and $m_3$ can be calculated using Eq.~\ref{eq:22} and mass square differences $\Delta m_{21}^2$ and 
$\Delta m_{32}^2$. That is
\begin{eqnarray}
\label{eq:24}
&& m_{1}=\sqrt{\Delta m_{21}^2}\sqrt{\frac{|\frac{m_{1}}{m_{2}}|^2} {|1-|\frac{m_{1}}{m_{2}}|^2|}}, \nonumber \\ 
&& m_{2}=\sqrt{|\Delta m_{32}^2|}\sqrt{\frac{1} {||\frac{m_{3}}{m_{2}}|^2-1|}}\,, \nonumber \\
&& m_{3}=\sqrt{|\Delta m_{32}^2|}\sqrt{\frac{1} {|1-|\frac{m_{2}}{m_{3}}|^2|}}\,.
 \end{eqnarray}

We can now explore whether or not the chosen texture of the neutrino mass matrix is empirically acceptable. We can construct the mass matrix
by using the allowed values of the experimental input parameters such as the mixing angles, mass squared differences and ratio $r$ 
and test whether or not the other experimental constraints are respected. We now proceed to discuss our results.

\section{Results and discussion}
\label{section:4}
Our main aim is to study the phenomenological implication of two minor zeros in the neutrino mass matrix on the total neutrino mass, the 
effective
Majorana mass term, the electron anti-neutrino mass and Majorana CP violating phases. From Eq.~\ref{eq:23}, it is clear that the CP violating
Majorana phases $\alpha$ and $\beta$ depend on $\theta$ and $\phi$. Similarly, from Eq.~\ref{eq:24}, it is clear that neutrino mass $m_{i}$ 
depends not only on $\theta$, $\phi$ but also on the mass squared differences $\Delta m_{21}^2$ and $\Delta m_{32}^2$. Moreover, all the 
neutrino oscillation parameters also depend only on the value of $\theta$ and $\phi$.
We first perform a $\chi^2$ analysis to find the best fit values of our model parameters $\theta$ and $\phi$ and test the validity of our
model. We define the $\chi^2$ as follows:
\begin{equation}
\label{chi2}
  \chi^2=\sum\limits_{i=1}^{3} \frac{\Big(\theta_{i}^{cal}- \theta_{i}^{exp}\Big)^2}{(\sigma_{i}^{exp})^2}+\sum\limits_{j=21,32} 
\frac{\Big(\Delta m_{j}^{cal}- \Delta m_{j}^{exp}\Big)^2}{(\sigma_{j}^{exp})^2}\,,
 \end{equation}
where $\theta_i=(\theta_{12},\theta_{13},\theta_{23})$ and $\Delta m_{j}= (\Delta m_{21}^2,\,\Delta m_{32}^2)$. Here $\theta_{i}^{cal}$ and
$\Delta m_{j}^{cal} $ represent the calculated value of $\theta_{i}$ and $\Delta m_{j}$, respectively, whereas, $\theta_{i}^{exp}$ and
$\Delta m_{j}^{exp} $ are the measured central values of $\theta_{i}$ and $\Delta m_{j}$, respectively. The $\theta_{i}^{cal}$ and
$\Delta m_{j}^{cal}$ depend on two unknown model parameters, namely $\theta$ and $\phi$. The $\sigma_{i}^{exp}$ and $\sigma_{j}^{exp}$ are the
uncertainties corresponding to the measured value of $\theta_{i}$ and $\Delta m_{j}$ respectively. The central values and the corresponding
uncertainties in each parameter, obtained from NuFIT~\cite{Esteban:2020cvm}, are reported in Table.~\ref{tab:2}.
Besides the best fit values of $\theta$ and $\phi$, the $\chi^2$ analysis also will return the best fit values of the neutrino
oscillation parameters such as the three mixing angles and the two mass squared differences for each class of two minor zero patterns. 
Moreover, we use the $3\sigma$ allowed range of $r$ to check the validity of our model. 
\begin{table}[htbp]
\begin{tabular}{|c|c|c|c|c|}
 \hline
 \hline
parameter & Normal ordering(best fit) & inverted ordering ($\Delta \chi^2=7.1)$\\
&bfp$\pm 1\sigma$ \hspace{1.5cm} $3\sigma$ ranges&bfp$\pm 1\sigma$ \hspace{1.5cm} $3\sigma$ ranges\\
\hline
$\theta_{12}^\circ
$&$33.44^{+0.77}_{-0.74}$ \hspace{1.5cm} 31.27$\rightarrow $ 35.86&$33.45^{+0.77}_{-0.74}$\hspace{1.5cm} 31.27$\rightarrow $ 35.87\\
\hline
$\theta_{23}^\circ$&$49.2^{+1.0}_{-1.3}$ \hspace{1.5cm} 39.5$\rightarrow $ 52.0&$49.5^{+1.0}_{-1.2}$\hspace{1.5cm} 39.8$\rightarrow $ 52.1\\
\hline
$\theta_{13}^\circ$&$8.57^{+0.13}_{-0.12}$ \hspace{1.5cm} 8.20$\rightarrow $ 8.97&$8.60^{+0.12}_{-0.12}$ \hspace{1.5cm} 8.24$\rightarrow $ 8.98\\
\hline
$\delta^\circ$&$194^{+52}_{-25}$ \hspace{1.5cm} 105$\rightarrow $ 405&$287^{+27}_{-32}$ \hspace{1.5cm} 192$\rightarrow $ 361\\
\hline
$\frac{\Delta m^2_{21}}{10^{-5}eV^2}$&$7.42^{+0.21}_{-0.20}$ \hspace{1.5cm} 6.82$\rightarrow $ 8.04&$7.42^{+0.21}_{-0.20}$ \hspace{1.5cm} 6.82$\rightarrow $ 8.04\\
\hline
$\frac{\Delta m^2_{3l}}{10^{-3}eV^2}$&$+2.515^{+0.028}_{-0.028}$ \hspace{1.5cm} +2.431$\rightarrow $ +2.599&$-2.498^{+0.028}_{-0.029}$ \hspace{1.5cm} -2.584$\rightarrow $ -2.413\\
\hline
\hline
\end{tabular}
\caption{Neutrino oscillation parameters from NuFIT~\cite{Esteban:2020cvm}. }
\label{tab:2}
\end{table}

It should be noted that minimum $\chi^2$ alone is not sufficient to determine the best mass model because it does not provide any information
regarding the degree of fine tuning in the mass matrix elements that is needed to reproduce the experimental data. In order to clarify this
issue, we will present a quantitative analysis regarding the degree of fine tuning in the elements of the neutrino mass matrix. In
case of TM$_1$ and TM$_2$ mixing matrix, the
elements of the neutrino mass matrix depend on two unknown parameters $\theta$ and $\phi$. The dimensionless quantity $d_{FT}$ which measures
the amount of fine tuning in the neutrino mass matrix element is defined as the sum of the absolute values of the
ratios between each parameter and its error~\cite{Altarelli:2010at,Meloni:2012sx}. We define $d_{FT}$ as
\begin{equation}
 d_{FT}=\sum \Big| \frac{par_i}{err_i}\Big|\,,
\end{equation}
where $par_i$ is the best fit values of the parameters $\theta$ and $\phi$, respectively. The error $err_i$ for each parameter is calculated
from the shift in the best fit value that changes $\chi^2_{min}$ value by one unit while keeping other parameters fixed at their best fit
values. Also we define $d_{Data}$ as the ratio of sum of absolute values of each parameter and their error. Using the data from
Table.~\ref{tab:2}, we obtain $d_{Data}$ to be around $100$. The $d_{FT}$ parameter can provide a rough estimate of the degree of fine
tuning in the mass matrix elements because if the $d_{FT}$ value is large then there will be large difference in the $\chi^2$ for a small
change in the corresponding parameter. Hence a large value of $d_{FT}$ corresponds to a strong fine tuning of the mass matrix elements and
vice versa.

We now proceed to analyse all the two minor zero classes one by one.

\subsection{\textbf{Class: $A_1$}} 
Class $A_1$ corresponds to the minor zero for the $(3,3)$ and the $(3,2)$ elements of the neutrino mass matrix. The corresponding equations
satisfying two minor zero conditions can be written as  
\begin{eqnarray}
\label{eq:28}
 &&(M)_{ ee}(M)_{ \mu\mu}-(M)_{ e\mu}(M)_{ \mu e}=0\,,\nonumber \\ 
 &&(M)_{ ee}(M)_{ \mu\tau}-(M)_{ \mu e}(M)_{ e\tau}=0\,.
\end{eqnarray}
Using Eq.~\ref{eq:22}, the mass ratios for TM$_1$ can be expressed as
\begin{eqnarray}
\label{eq:29} 
&&\frac{m_1}{m_2}e^{-2i\alpha}=\frac{-\cos\theta}{2\sqrt{3}(\frac{1}{\sqrt{3}}\cos\theta+\frac{1}{\sqrt{2}}\sin\theta  e^{-i\phi})}\,,\nonumber \\ 
&&\frac{m_1}{m_3}e^{-2i\beta}=\frac{-\sin\theta}{2\sqrt{3}(\frac{1}{\sqrt{3}}\sin\theta-\frac{1}{\sqrt{2}}\cos\theta  e^{-i\phi})}\,,\nonumber \\ 
&&\frac{m_3}{m_2}e^{-2i(\alpha+\beta)}=\frac{\cos\theta(\frac{1}{\sqrt{3}}\sin\theta-\frac{1}{\sqrt{2}}\cos\theta  e^{-i\phi})}{\sin\theta (\frac{1}{\sqrt{3}}\cos\theta+\frac{1}{\sqrt{2}}\sin\theta  e^{-i\phi})}\,.
\end{eqnarray}
Similarly, for TM$_2$ mixing matrix, the mass ratios can be expressed as
\begin{eqnarray} 
\label{eq:30}
&&\frac{m_1}{m_2}e^{-2i\alpha}=\frac{-(\cos\theta+\sqrt{3}\sin\theta  e^{i\phi})}{2\cos\theta}\,,\nonumber \\ 
&&\frac{m_1}{m_3}e^{-2i\beta}=\frac{\sin\theta(\frac{1}{\sqrt{6}}\cos\theta+\frac{1}{\sqrt{2}}\sin\theta  e^{i\phi})}{\cos\theta (\frac{1}{\sqrt{6}}\sin\theta-\frac{1}{\sqrt{2}}\cos\theta  e^{i\phi})}\,,\nonumber \\ 
&&\frac{m_3}{m_2}e^{-2i(\alpha+\beta)}=\frac{-\sin\theta+\sqrt{3}\cos\theta  e^{i\phi}}{2\sin\theta}\,.
\end{eqnarray}
The $\chi_{min}^2$ value and the $d_{FT}$ parameter for Class $A_1$ are listed in Table.~\ref{tab:3}. It is evident that the fine tuning 
parameter $d_{FT}$ is quite large in case of TM$_{1}$ mixing matrix compared to TM$_{2}$ mixing matrix for this class. Hence the degree of 
fine tuning in the elements of neutrino mass matrix is quite strong in case of TM$_{1}$ mixing matrix. The corresponding best fit values of 
our model parameters $\theta$ and $\phi$ along with all the neutrino oscillation parameters namely, the three mixing angles~$(\theta_{13},\, 
\theta_{23},\, \theta_{12})$, two mass squared differences~$(\Delta m^2_{21},\, \Delta m^2_{32})$, the Jarsklog
invariant $J$ and Dirac CP violating phase $\delta$ obtained for this class are reported in Table.~\ref{tab:3}.
It is evident from Eq.~\ref{eq:7} and Eq.~\ref{eq:14} that $\theta_{23}$ is invariant under the transformation 
$\phi \to (2\pi - \phi)$, hence we get two best fit values of $\phi$. Similarly, we get two values of $J$ and $\delta$ corresponding to two 
best fit values of $\phi$. The best fit values of the mixing angles $\theta_{12}$, $\theta_{13}$ and the mass squared differences 
$\Delta m^2_{21}$, $\Delta m^2_{32}$ obtained for this class are compatible with the experimentally measured values reported in 
Table.~\ref{tab:2}. It is observed that, in case of TM$_2$ mixing matrix, the best fit value of $\theta_{23}$ is compatible with the 
experimentally measured value but for TM$_1$ mixing matrix the best fit value deviates significantly from the experimentally measured value.
Moreover, in case of TM$_2$ mixing matrix, the best fit value of $\theta_{12}$ deviates from the measured value of $\theta_{12}$ at more
than $2\sigma$ significance. 

\begin{table}[htbp]
\begin{tabular}{|c|c|c|c|c|c|c|c|c|c|c|c|}
 \hline
 Mixing& $\chi^2_{min}$ &$d_{FT}$&$\phi^\circ$&$\theta^\circ $& $\theta_{12}^\circ $&$\theta_{13}^\circ $&$\theta_{23}^\circ $&$J$&$\delta^\circ$&$\Delta m^2_{21}\,{(10^{-5}\rm eV^2)}$ & $\Delta m^2_{32}\,{(10^{-3}\rm eV^2)}$\\
 matrix&&&&&&&&&&&\\
\hline
TM$_1$& $33.37$ &$1.24\times$$10^2$ &$ 101.67,258.32$&$14.67$&$34.37$&$8.41$&$42.62$&$\pm3.26\times 10^{-2}$& $79.32,280.68$&7.47&2.50\\
\hline
TM$_2$& $20.61$ &$1.25$ &$ 37.59,322.41$&$10.03$&$35.68$&$8.17$&$49.61$&$\pm2.01\times 10^{-2}$& $38.17,321.83$&7.55&2.48\\
\hline
\hline
\end{tabular}
\caption{$\chi_{min}^2$, $d_{FT}$, best fit values of $\phi^\circ$, $\theta^\circ $, $\theta_{12}^\circ $, $\theta_{13}^\circ $, $\theta_{23}^\circ $, J, 
$\delta^\circ$ , $\Delta m^2_{21}\,{(10^{-5}\rm eV^2)}$ and $\Delta m^2_{32}\,{(10^{-3}\rm eV^2)}$ for TM$_1$ and TM$_2$ mixing matrix for Class $A_1$.}
\label{tab:3}
\end{table}

\begin{figure}[htbp!]
\begin{subfigure}{0.32\textwidth}
\includegraphics[width=\textwidth]{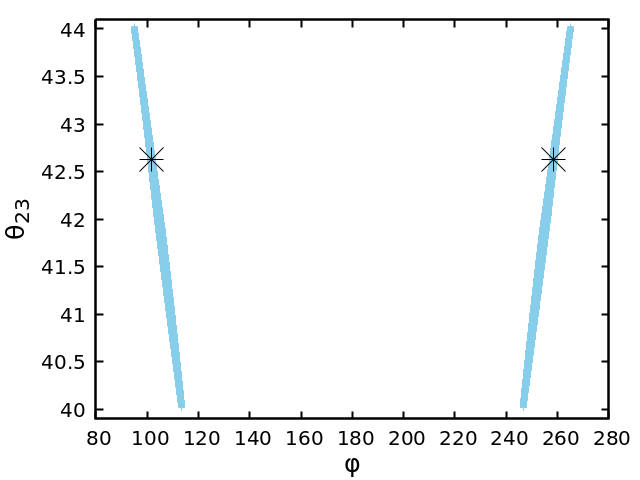}
\caption{}
\label{fig:TM1cla1ff1}
\end{subfigure}
\begin{subfigure}{0.32\textwidth}
\includegraphics[width=\textwidth]{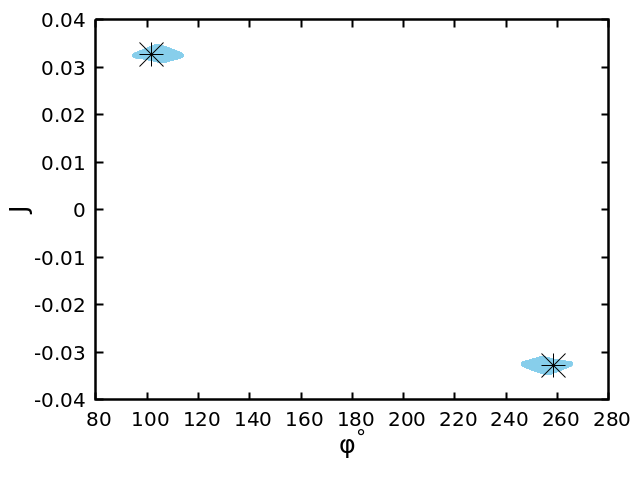}
\caption{}
\label{fig:TM1cla1ff2}
\end{subfigure}
\begin{subfigure}{0.32\textwidth}
\includegraphics[width=\textwidth]{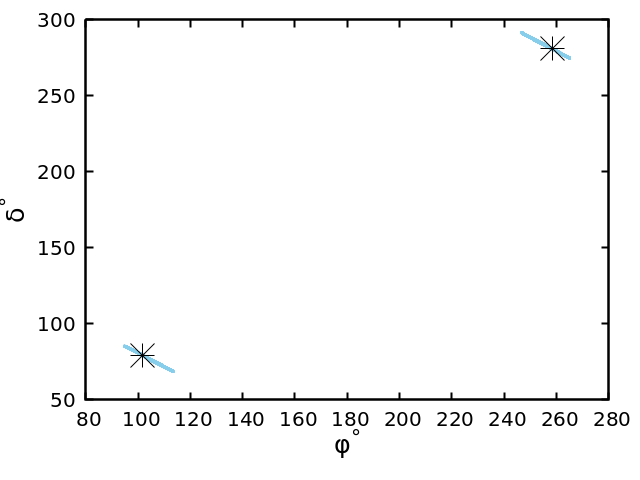}
\caption{}
\label{fig:TM1cla1ff3}
\end{subfigure}
\captionsetup{justification=raggedright,singlelinecheck=false}
\caption{ (a) Variation of $\theta_{23}$ as a function of $\phi$, (b) variation of $J$ as a function of $\phi$, and (c) variation of $\delta$ 
as a function of $\phi$ for TM$_1$ mixing matrix for Class $A_1$. The '$*$' mark in the figures represents the best fit value.}
\end{figure}

For the TM$_1$ mixing matrix, we use Eq.~\ref{eq:7} and vary $\theta_{13}$ within $3\sigma$ from the central value and obtain the $3\sigma$ 
allowed range of $\theta$ to be $(14.26^\circ - 15.64^\circ)$. Using
the allowed range of $\theta$ and imposing the additional constraint coming from $r$, we obtain the allowed ranges of $\theta_{12}$ and 
$\theta_{23}$ to be $(34.25^\circ - 34.42^\circ)$ and $(40.01^\circ - 44.02^\circ)$, respectively. It is clear that the value of $\theta_{23}$
obtained in this case lies in the lower octant, i.e, for the TM$_1$ mixing matrix, this pattern prefers the atmospheric mixing angle
to be smaller than $\pi/4$. We show the variation of $\theta_{23}$ as a function of the unknown parameter $\phi$ in Fig.~\ref{fig:TM1cla1ff1}.
The corresponding best fit value of $\theta_{23}$ is shown with '*' mark in Fig.~\ref{fig:TM1cla1ff1}. We show the variation of $J$ and 
$\delta$ as a function of $\phi$ in Fig.~\ref{fig:TM1cla1ff2} and Fig.~\ref{fig:TM1cla1ff3}, respectively. 
It is observed that the Jarlskog rephasing invariant $J$ and the Dirac CP violating phase $\delta$ are restricted to two regions.
We obtain the $3\sigma$ allowed 
ranges of $J$ and $\delta$ to be $[(-3.12\times 10^{-2}, -3.43\times 10^{-2}),\, (3.12\times 10^{-2}, 3.43\times 10^{-2})]$ and
$[(68.66, 85.48)^\circ,\,(274.51, 291.33)^\circ]$, respectively.

\begin{figure}[htbp]
\begin{subfigure}{0.32\textwidth}
\includegraphics[width=\textwidth]{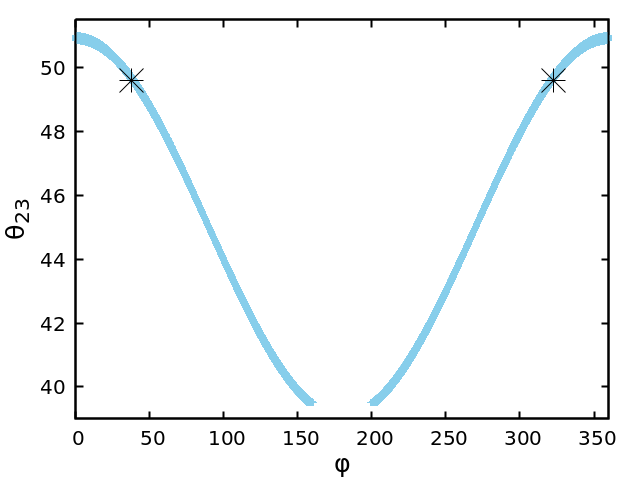}
\caption{}
\label{fig:TM2cla1ff1}
\end{subfigure}
\begin{subfigure}{0.32\textwidth}
\includegraphics[width=\textwidth]{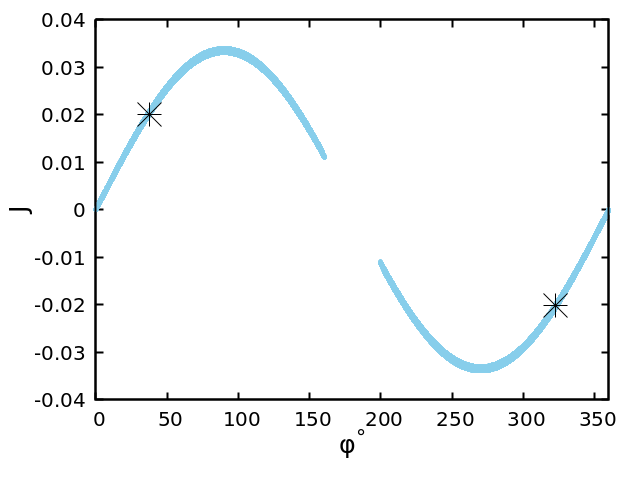}
\caption{}
\label{fig:TM2cla1ff2}
\end{subfigure}
\begin{subfigure}{0.32\textwidth}
\includegraphics[width=\textwidth]{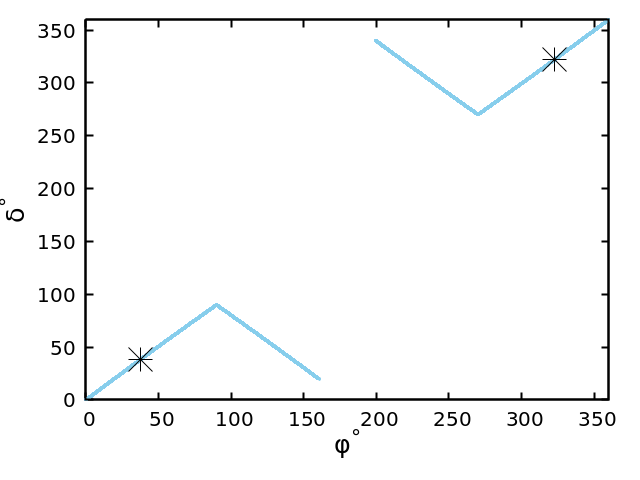}
\caption{}
\label{fig:TM2cla1ff3}
\end{subfigure}
\captionsetup{justification=raggedright,singlelinecheck=false}
\caption{ (a) Variation of $\theta_{23}$ as a function of $\phi$, (b) variation of $J$ as a function of $\phi$, and (c) variation of $\delta$ 
as a function of $\phi$ for TM$_2$ mixing matrix for Class $A_1$. The '$*$' mark in the figures represents the best fit value.}
\end{figure}

For the TM$_2$ mixing matrix, we use Eq.~\ref{eq:14} and obtain the $3\sigma$ allowed range of $\theta$ to be $(10.03^\circ - 10.99^\circ)$.
The corresponding allowed ranges of $\theta_{12}$ and $\theta_{23}$ are found to be $(35.68^\circ - 35.75^\circ)$ and 
$(39.00^\circ - 50.99^\circ)$, respectively. We show the variation of $\theta_{23}$ as a function of the unknown parameter $\phi$ in 
Fig.~\ref{fig:TM2cla1ff1}. We also show the variation of $J$ and  $\delta$ as a function of $\phi$ in Fig.~\ref{fig:TM2cla1ff2} and  
Fig.~\ref{fig:TM2cla1ff3}, respectively. The $3\sigma$ allowed ranges of $J$ and $\delta$ are found to be $[0,\pm 3.39\times 10^{-2}]$ and 
$[(0, 90)^\circ,\,(270, 360)^\circ]$, respectively.

\begin{figure}[htbp]
\begin{subfigure}{0.32\textwidth}
\includegraphics[width=\textwidth]{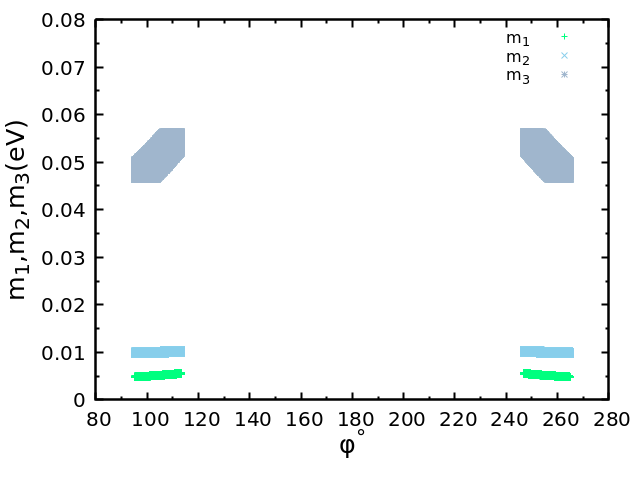}
\caption{}
\label{fig:3a}
\end{subfigure}
\begin{subfigure}{0.32\textwidth}
\includegraphics[width=\textwidth]{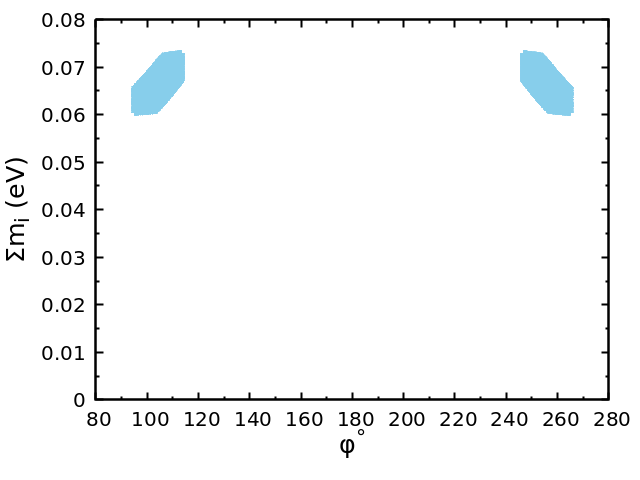}
\caption{}
\label{fig:3b}
\end{subfigure}
\begin{subfigure}{0.32\textwidth}
\includegraphics[width=\textwidth]{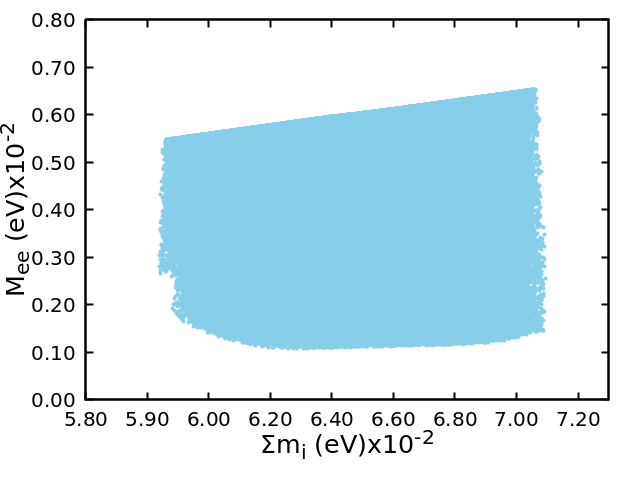}
\caption{}
\label{fig:3c}
\end{subfigure}
\begin{subfigure}{0.32\textwidth}
\includegraphics[width=\textwidth]{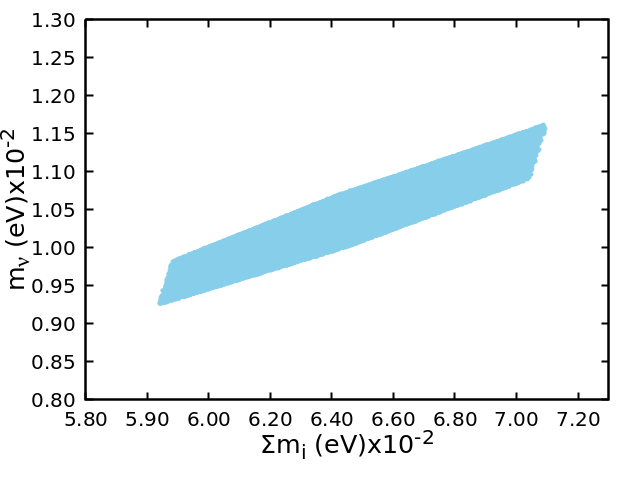}
\caption{}
\label{fig:3d}
\end{subfigure}
\begin{subfigure}{0.32\textwidth}
\includegraphics[width=\textwidth]{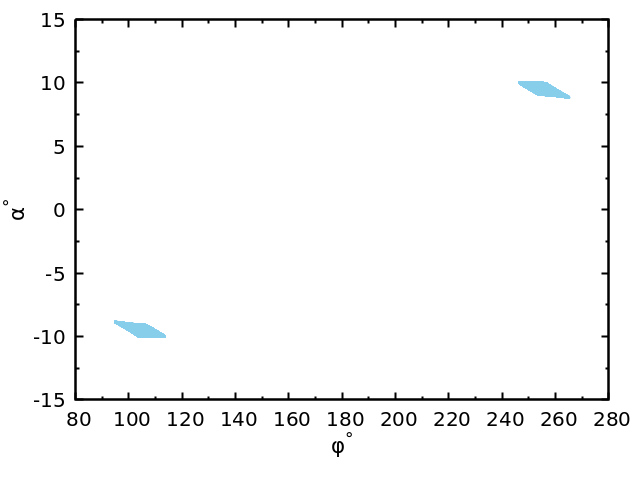}
\caption{}
\label{fig:3e}
\end{subfigure}
\begin{subfigure}{0.32\textwidth}
\includegraphics[width=\textwidth]{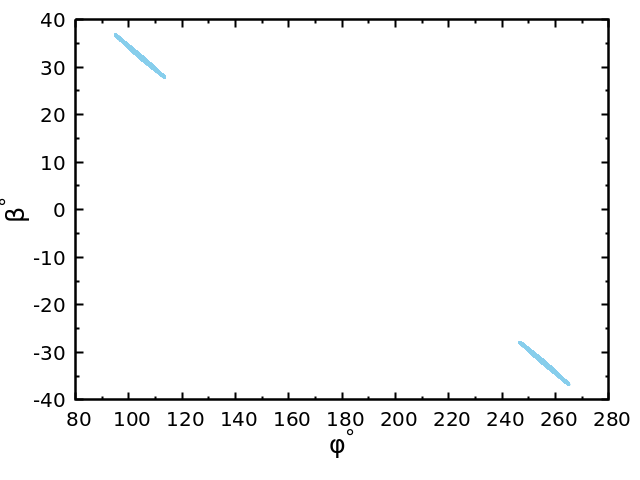}
\caption{}
\label{fig:3f}
\end{subfigure}
\captionsetup{justification=raggedright,singlelinecheck=false}
\caption{ (a) Variation of $m_1$, $m_2$, $m_3$ as a function of $\phi$, (b) variation of $\sum m_i$ as a function of $\phi$, (c) correlation 
between $\sum m_i$ and $M_{ee}$, (d) correlation between $\sum m_i$ and $m_{\nu}$,  (e) variation of $\alpha$ as a function of $\phi$, and (f)
variation of $\beta$ as a function of $\phi$ for TM$_1$ mixing matrix for Class $A_1$.}
\label{fig:3}
\end{figure}

\begin{figure}[htbp]
\begin{subfigure}{0.32\textwidth}
\includegraphics[width=\textwidth]{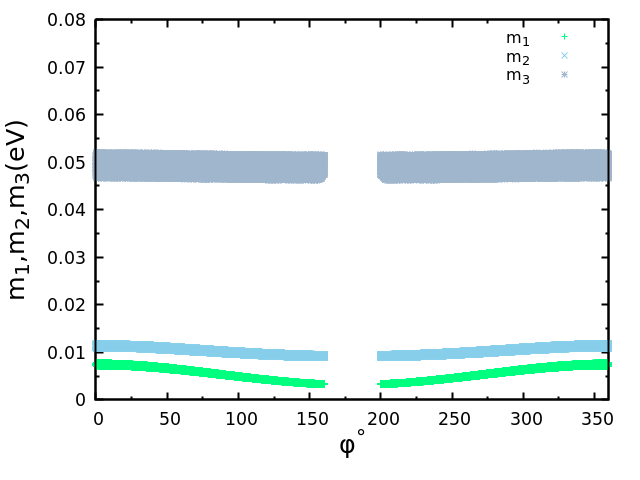}
\caption{}
\label{fig:4a}
\end{subfigure}
\hfill
\begin{subfigure}{0.32\textwidth}
\includegraphics[width=\textwidth]{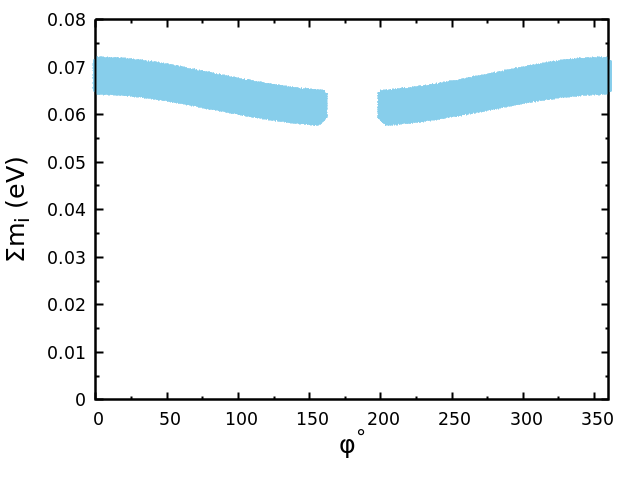}
\caption{}
\label{fig:4b}
\end{subfigure}
\hfill
\begin{subfigure}{0.32\textwidth}
\includegraphics[width=\textwidth]{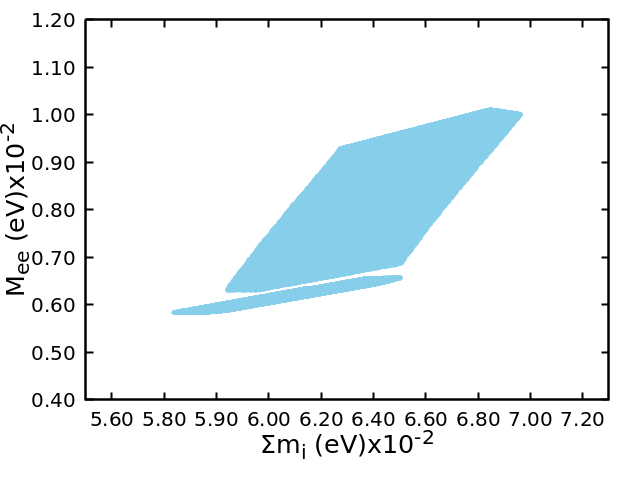}
\caption{}
\label{fig:4c}
\end{subfigure}
\hfill
\begin{subfigure}{0.32\textwidth}
\includegraphics[width=\textwidth]{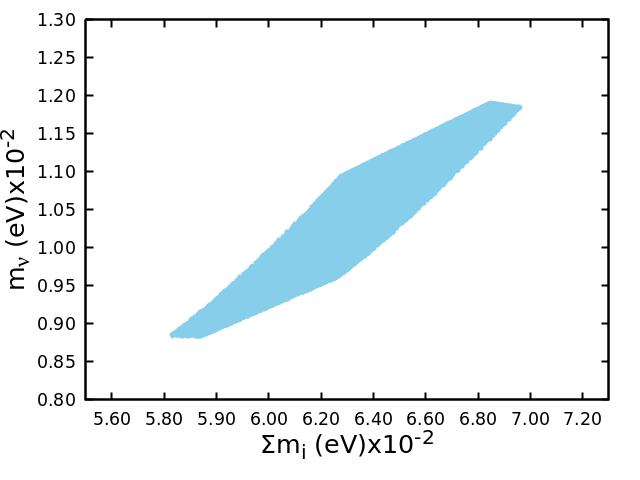}
\caption{}
\label{fig:4d}
\end{subfigure}
\hfill
\begin{subfigure}{0.32\textwidth}
\includegraphics[width=\textwidth]{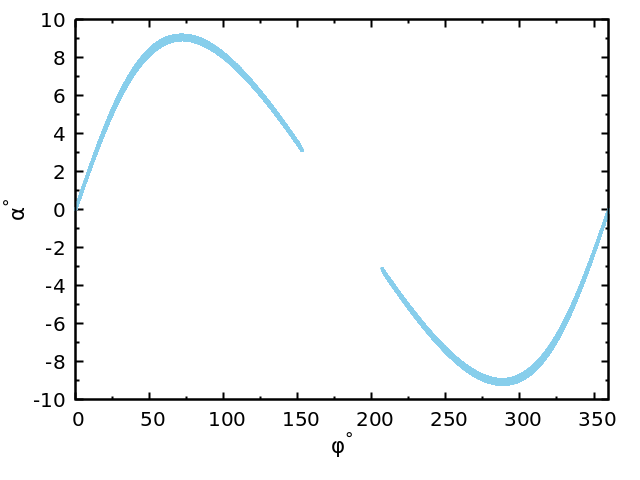}
\caption{}
\label{fig:4e}
\end{subfigure}
\hfill
\begin{subfigure}{0.32\textwidth}
\includegraphics[width=\textwidth]{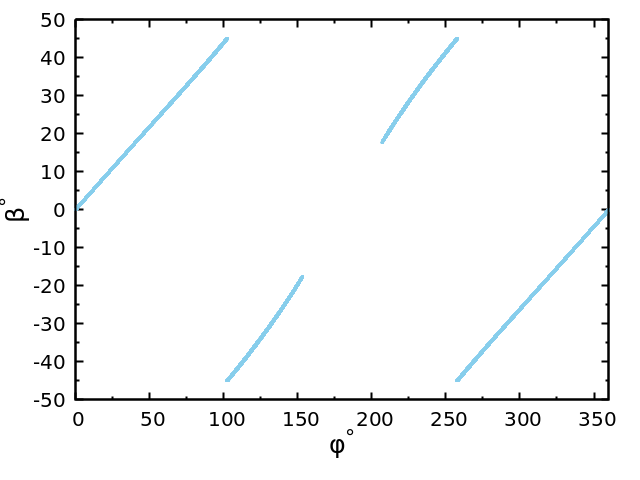}
\caption{}
\label{fig:4f}
\end{subfigure}
\captionsetup{justification=raggedright,singlelinecheck=false}
\caption{ (a) Variation of $m_1$, $m_2$, $m_3$ as a function of $\phi$, (b) variation of $\sum m_i$ as a function of $\phi$, (c) correlation 
between $\sum m_i$ and $M_{ee}$, (d) correlation between $\sum m_i$ and $m_{\nu}$,  (e) variation of $\alpha$ as a function of $\phi$, and (f)
variation of $\beta$ as a function of $\phi$ for TM$_2$ mixing matrix for Class $A_1$.}
\label{fig:4}
\end{figure}

We show the variation of neutrino masses $m_1$, $m_2$ and $m_3$ as a function of $\phi$ in Fig~\ref{fig:3a} and Fig.~\ref{fig:4a} for TM$_1$ 
and TM$_2$ mixing matrix, respectively. It shows normal mass ordering for both TM$_1$ and TM$_2$ mixing matrix. In Fig~\ref{fig:3b} and 
Fig.~\ref{fig:4b}, we show the variation of $\sum m_i$ as a function of $\phi$. The correlation of $M_{ee}$ and $\sum m_i$ for TM$_1$ and 
TM$_2$ mixing matrix are shown in Fig.~\ref{fig:3c} and Fig.~\ref{fig:4c}, respectively. In Fig.~\ref{fig:3d} and Fig.~\ref{fig:4d}, we have 
shown the correlation of $m_{\nu}$ with $\sum m_i$ for TM$_1$ and TM$_2$ mixing matrix, respectively. The variation of Majorana phases 
$\alpha$ and $\beta$ as a function of $\phi$ is shown in Fig.~\ref{fig:3e} and Fig.~\ref{fig:3f} for TM$_1$ mixing matrix and in 
Fig.~\ref{fig:4e} and Fig.~\ref{fig:4f} for TM$_2$ mixing matrix, respectively.   

The best fit values and the corresponding $3\sigma$ allowed ranges of the absolute neutrino mass scale, the effective Majorana neutrino mass, 
the effective electron anti-neutrino mass and CP violating phases $\alpha$ and $\beta$ are listed in Table.~\ref{tab:4}. It is observed that
the CP violating Majorana phases $\alpha$ and $\beta$ are restricted to two regions. For TM$_1$ mixing matrix, the best fit values of 
$\alpha$ and $\beta$ are obtained to be $\pm 9.28$ and $\pm 33.48$, respectively. Similarly, for the TM$_2$ mixing matrix, the best fit values
of $\alpha$ and $\beta$ are found to be $\pm 6.62$ and $\pm 16.46$, respectively. The upper bound of $M_{ee}$ obtained for this class is of 
$\mathcal O(10^{-2})$ and
is within the sensitivity reach of neutrinoless double beta decay. The upper bound on the effective electron anti-neutrino mass 
$m_{\nu}<0.012\,{\rm eV}$ is beyond the reach of current $\beta$ decay experiments. 
\begin{table}[htbp!]
\begin{tabular}{|c|c|c|c|c|c|c|}
 \hline
 Mixing&Values& $\sum m_i\,{(\rm eV)}$ &$M_{ee}\,{(\rm eV)}$&$m_{\nu}\,{(\rm eV)}$&$\alpha^\circ$&$\beta^\circ$\\
 matrix&& &&&&\\
\hline
\multirow{2}{*}{TM$_1$}&Best fit& $0.066$ & $0.006$&$0.010$&$\pm 9.28$& $\pm 33.48$\\
\cline{2-7}
&$3\sigma$ Range & $[0.060, 0.073]$ & $[0.001, 0.007]$&$[0.009, 0.011]$&$[(-10.03,-8.84),\,(8.84, 10.03)]$& $[(-36.79,-27.82),\,(27.82, 36.79)]$\\
\hline
\multirow{2}{*}{TM$_2$}&Best fit& $0.069$ & $0.009$&$0.011$&$\pm 6.62$&$\pm 16.46$\\ 
\cline{2-7}
&$3\sigma$ Range& $[0.058, 0.071]$ & $[0.005, 0.010]$&$[0.009, 0.012]$&$[0,\pm 9.18]$&$[0,\pm 44.99]$\\
\hline
\hline
\end{tabular}
\caption{Best fit and $3\sigma$ allowed range of $\sum m_{i}(\rm eV)$, $M_{ee}(\rm eV)$, $m_{\nu}(\rm eV)$, $\alpha^\circ$ and $\beta^\circ$ 
for Class $A_1$.}
\label{tab:4}
\end{table}

\subsection{\textbf{Class: $A_2$}} 
For Class $A_2$, the minors corresponding to the $(2,2)$ and $(3,2)$ elements of the neutrino mass matrix are zero. The minor zero conditions
for this class can be written in terms of the neutrino mass matrix elements as
\begin{eqnarray} 
\label{eq:31}
 &&(M)_{ ee}(M)_{ \tau\tau}-(M)_{ e\tau}(M)_{ \tau e}=0\,,\nonumber \\ 
 &&(M)_{ ee}(M)_{ \mu\tau}-(M)_{ \mu e}(M)_{ e\tau}=0\,.
\end{eqnarray}
The mass ratios for TM$_1$ and TM$_2$ mixing matrix can be expressed as
\begin{eqnarray}
\label{eq:32}
&&\frac{m_1}{m_2}e^{-2i\alpha}=\frac{-\cos\theta}{2\sqrt{3}(\frac{1}{\sqrt{3}}\cos\theta-\frac{1}{\sqrt{2}}\sin\theta  e^{-i\phi})}\,,\nonumber \\ 
&&\frac{m_1}{m_3}e^{-2i\beta}=\frac{-\sin\theta}{2\sqrt{3}(\frac{1}{\sqrt{3}}\sin\theta+\frac{1}{\sqrt{2}}\cos\theta  e^{-i\phi})}\,,\nonumber \\ 
&&\frac{m_3}{m_2}e^{-2i(\alpha+\beta)}=\frac{\cos\theta(\frac{1}{\sqrt{3}}\sin\theta+\frac{1}{\sqrt{2}}\cos\theta  e^{-i\phi})}{\sin\theta (\frac{1}{\sqrt{3}}\cos\theta-\frac{1}{\sqrt{2}}\sin\theta  e^{-i\phi})}\,.
\end{eqnarray}
and
\begin{eqnarray}
\label{eq:33}
&&\frac{m_1}{m_2}e^{-2i\alpha}=\frac{-\cos\theta+\sqrt{3}\sin\theta  e^{i\phi}}{2\cos\theta}\,,\nonumber \\ 
&&\frac{m_1}{m_3}e^{-2i\beta}=\frac{\sin\theta(\frac{1}{\sqrt{6}}\cos\theta-\frac{1}{\sqrt{2}}\sin\theta  e^{i\phi})}{\cos\theta (\frac{1}{\sqrt{6}}\sin\theta+\frac{1}{\sqrt{2}}\cos\theta  e^{i\phi})}\,,\nonumber \\ 
&&\frac{m_3}{m_2}e^{-2i(\alpha+\beta)}=\frac{-(\sin\theta+\sqrt{3}\cos\theta  e^{i\phi})}{2\sin\theta}\,.
\end{eqnarray}
We report the values of $\chi_{min}^2$ and $d_{FT}$ for Class $A_2$ in Table.~\ref{tab:5}. The fine tuning parameter $d_{FT}$ is found to be 
quite large in case of TM$_{1}$ mixing matrix. So, for this class, the degree of fine tuning of the elements of neutrino mass matrix is 
strong in case of TM$_{1}$ mixing matrix although $\chi_{min}^2$ value obtained is quite small. Corresponding best fit values of our model 
parameters $\theta$ and $\phi$ along with all the neutrino oscillation parameters namely, the three mixing
angles~$(\theta_{13},\, \theta_{23},\, \theta_{12})$, two mass squared differences~$(\Delta m^2_{21},\, \Delta m^2_{32})$, the Jarsklog
invariant $J$ and Dirac CP violating phase $\delta$ obtained for this class are reported in Table.~\ref{tab:5}.
The best fit values of the mixing angles $\theta_{12}$, $\theta_{13}$, $\theta_{23}$ and the mass squared differences 
$\Delta m^2_{21}$, $\Delta m^2_{32}$ obtained are compatible with the experimentally measured values. It should, however, be noted that the
best fit value of $\theta_{12}$ for the TM$_2$ mixing matrix differs from the experimentally measured central value at more than $2\sigma$ 
significance. This is quite a generic feature of TM$_2$ mixing matrix.

\begin{table}[htbp]
\begin{tabular}{|c|c|c|c|c|c|c|c|c|c|c|c|}
 \hline
 Mixing& $\chi^2_{min}$ &$d_{FT}$&$\phi^\circ$&$\theta^\circ $& $\theta_{12}^\circ $&$\theta_{13}^\circ $&$\theta_{23}^\circ $&$J$&$\delta^\circ$&$\Delta m^2_{21}\,{(10^{-5}\rm eV^2)}$ & $\Delta m^2_{32}\,{(10^{-3}\rm eV^2)}$\\
 matrix&&&&&&&&&&&\\
\hline
TM$_1$& $1.65$ &$5.63\times$$10^2$&$ 72.55,287.45$&$15.02$&$34.33$&$8.60$&$48.60$&$\pm3.25\times 10^{-2}$& $74.07,285.93$&7.42&2.59\\
\hline
TM$_2$& $23.28$ &$1.30$ &$ 46.68,313.31$&$10.00$&$35.67$&$8.14$&$48.96$&$\pm2.39\times 10^{-2}$& $47.29,312.71$&7.62&2.47\\
\hline
\hline
\end{tabular}
\caption{$\chi_{min}^2$, $d_{FT}$, best fit values of $\phi^\circ$, $\theta^\circ $, $\theta_{12}^\circ $, $\theta_{13}^\circ $, $\theta_{23}^\circ $, $J$, 
$\delta^\circ$ , $\Delta m^2_{21}\,{(10^{-5}\rm eV^2)}$ and $\Delta m^2_{32}\,{(10^{-3}\rm eV^2)}$ for Class $A_2$.}
\label{tab:5}
\end{table}

\begin{figure}[htbp]
\begin{subfigure}{0.32\textwidth}
\includegraphics[width=\textwidth]{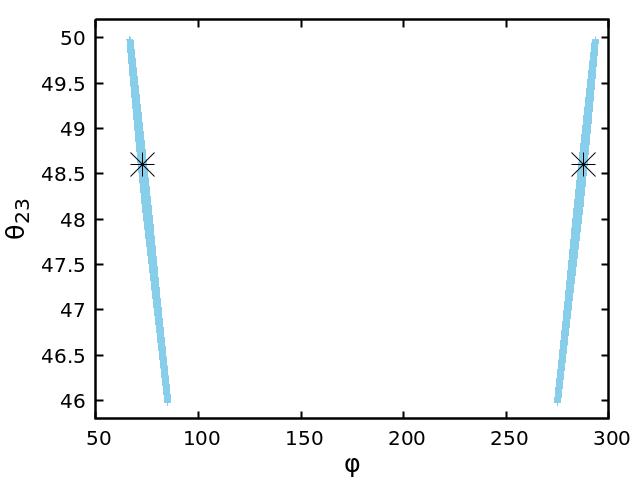}
\caption{}
\label{fig:TM1cla2ff1}
\end{subfigure}
\begin{subfigure}{0.32\textwidth}
\includegraphics[width=\textwidth]{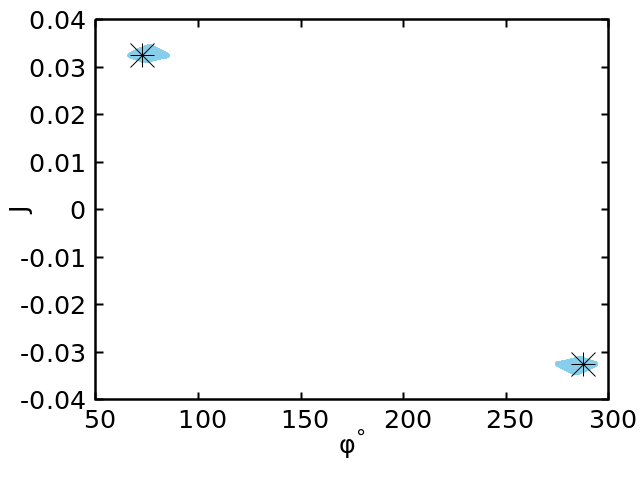}
\caption{}
\label{fig:TM1cla2ff2}
\end{subfigure}
\begin{subfigure}{0.32\textwidth}
\includegraphics[width=\textwidth]{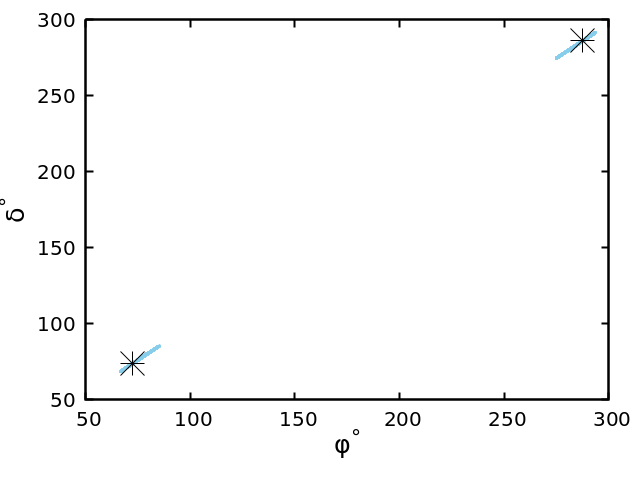}
\caption{}
\label{fig:TM1cla2ff3}
\end{subfigure}
\captionsetup{justification=raggedright,singlelinecheck=false}
\caption{ (a) Variation of $\theta_{23}$ as a function of $\phi$, (b) variation of $J$ as a function of $\phi$, and (c) variation of $\delta$ 
as a function of $\phi$ for TM$_1$ mixing matrix for Class $A_2$. The '$*$' mark in the figures represents the best fit value.}
\end{figure}

For the TM$_1$ mixing matrix, the allowed range of $\theta$, obtained by using the $3\sigma$ experimental range of $\theta_{13}$ reported in 
Table.~\ref{tab:2}, is $(14.26^\circ - 15.64^\circ)$. Using the allowed range of $\theta$, we obtain the allowed ranges of $\theta_{12}$ and
$\theta_{23}$ to be $(34.25^\circ - 34.42^\circ)$ and $(45.97^\circ - 49.98^\circ)$, respectively. We also use the constraint coming from
$r$ to constrain the allowed parameter space. It is clear
that the value of $\theta_{23}$ obtained in this case lies in the upper octant, i.e, for the TM$_1$ mixing matrix, this pattern prefers
the atmospheric mixing angle to be higher than $\pi/4$. We show the variation of $J$ and $\delta$ as a function of $\phi$ in 
Fig.~\ref{fig:TM1cla2ff2} and Fig.~\ref{fig:TM1cla2ff3}, respectively. We obtain the $3\sigma$ allowed ranges of $J$ and $\delta$ to be
$[(-3.12\times 10^{-2}, -3.43\times 10^{-2}),\, (3.12\times 10^{-2}, 3.43\times 10^{-2})]$ and 
$[(68.66, 85.48)^\circ,\,(274.51, 291.33)^\circ]$, respectively. Similar to Class $A_1$, it is observed that the Jarlskog rephasing
invariant $J$ and the Dirac CP violating phase $\delta$ are restricted to two regions.

\begin{figure}[htbp]
\begin{subfigure}{0.32\textwidth}
\includegraphics[width=\textwidth]{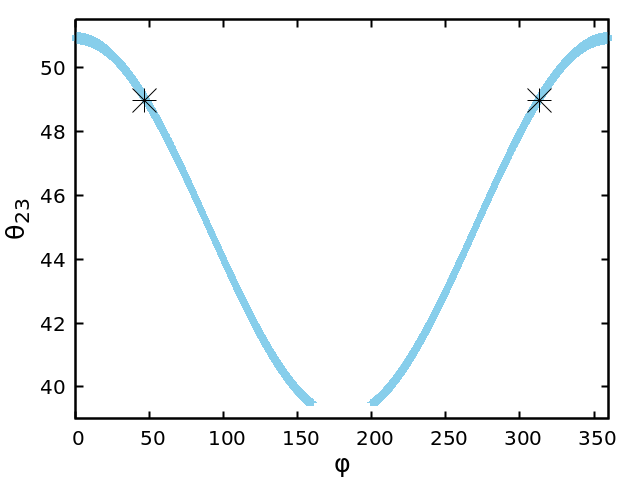}
\caption{}
\label{fig:TM2cla2ff1}
\end{subfigure}
\begin{subfigure}{0.32\textwidth}
\includegraphics[width=\textwidth]{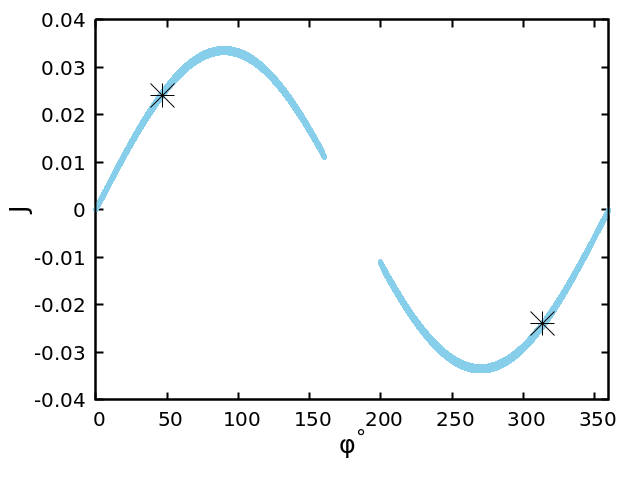}
\caption{}
\label{fig:TM2cla2ff2}
\end{subfigure}
\begin{subfigure}{0.32\textwidth}
\includegraphics[width=\textwidth]{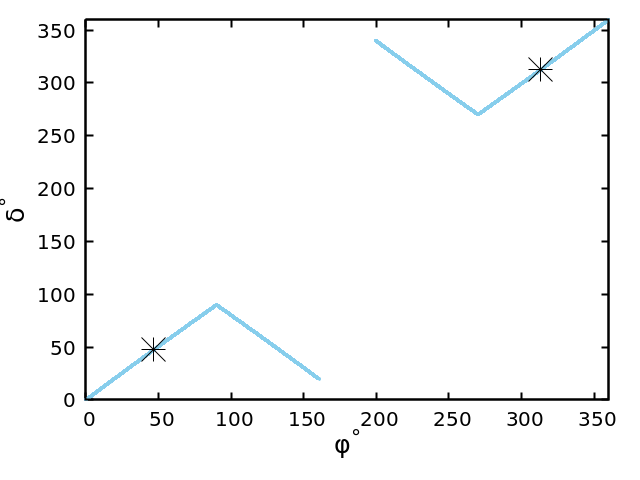}
\caption{}
\label{fig:TM2cla2ff3}
\end{subfigure}
\captionsetup{justification=raggedright,singlelinecheck=false}
\caption{ (a) Variation of $\theta_{23}$ as a function of $\phi$, (b) variation of $J$ as a function of $\phi$, and (c) variation of $\delta$ 
as a function of $\phi$ for TM$_2$ mixing matrix for Class $A_2$. The '$*$' mark in the figures represents the best fit value.}
\end{figure}

For the TM$_2$ mixing matrix, the $3\sigma$ allowed range of $\theta$ is found to be $(10.03^\circ - 10.99^\circ)$. Corresponding $3\sigma$ 
allowed range of $\theta_{12}$ and $\theta_{23}$ are $(35.68^\circ - 35.75^\circ)$ and $(39.00^\circ - 50.99^\circ)$, respectively. We show 
the variation of $\theta_{23}$ as a function of the unknown parameter $\phi$ in Fig.~\ref{fig:TM2cla2ff1}. The $3\sigma$ allowed ranges of $J$
and $\delta$ are found to be $[0,\pm 3.39\times 10^{-2}]$ and $[(0, 90)^\circ,\,(270, 360)^\circ]$, respectively. We also show the variation 
of $J$ and $\delta$ as a function of $\phi$ in Fig.~\ref{fig:TM2cla2ff2} and Fig.~\ref{fig:TM2cla2ff3}, respectively.

\begin{figure}[htbp]
\begin{subfigure}{0.32\textwidth}
\includegraphics[width=\textwidth]{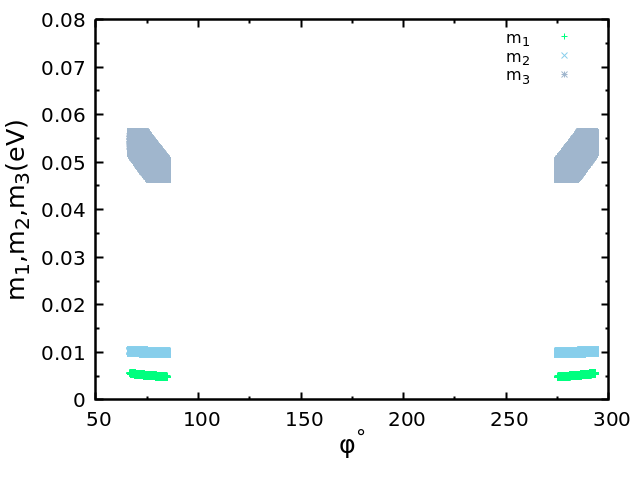}
\caption{}
\label{fig:5a}
\end{subfigure}
\begin{subfigure}{0.32\textwidth}
\includegraphics[width=\textwidth]{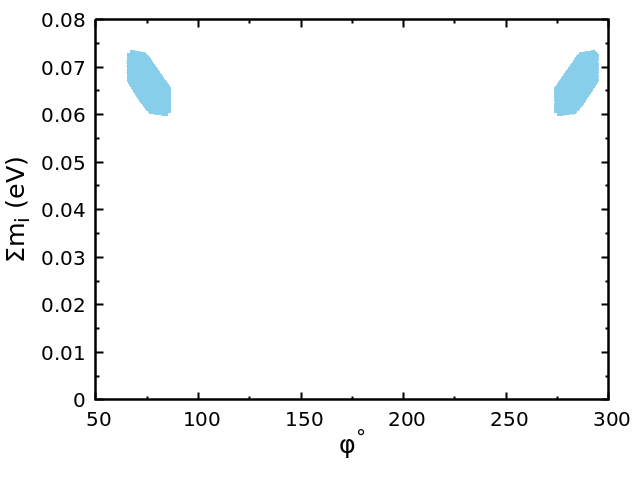}
\caption{}
\label{fig:5b}
\end{subfigure}
\begin{subfigure}{0.32\textwidth}
\includegraphics[width=\textwidth]{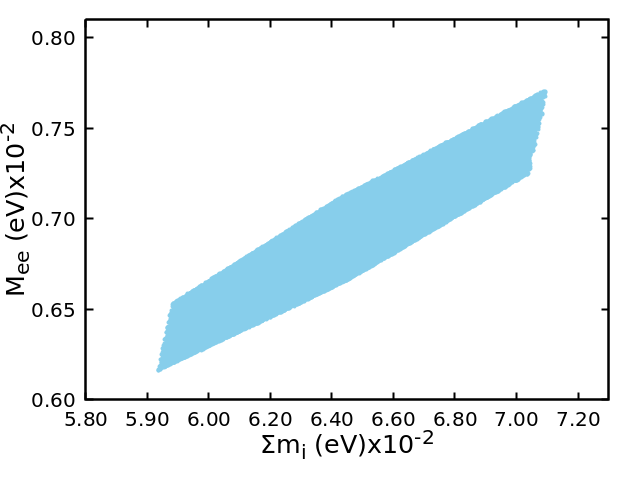}
\caption{}
\label{fig:5c}
\end{subfigure}
\begin{subfigure}{0.32\textwidth}
\includegraphics[width=\textwidth]{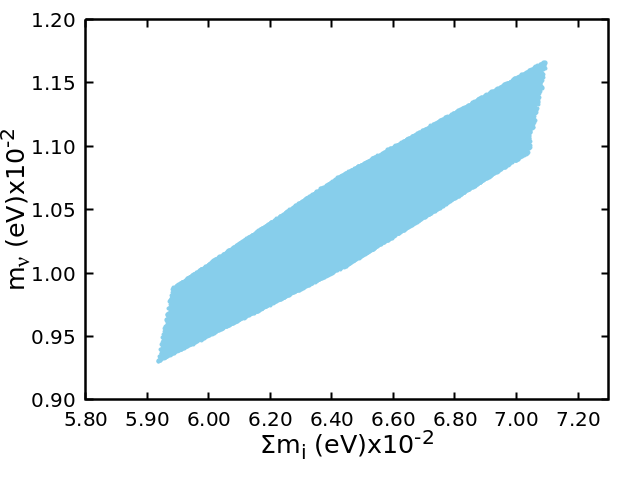}
\caption{}
\label{fig:5d}
\end{subfigure}
\begin{subfigure}{0.32\textwidth}
\includegraphics[width=\textwidth]{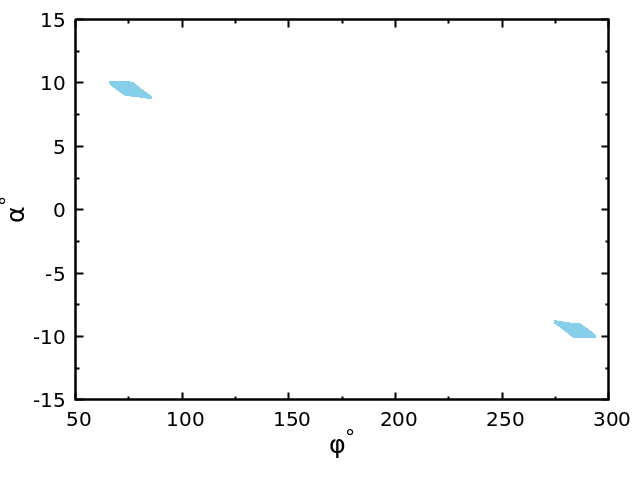}
\caption{}
\label{fig:5e}
\end{subfigure}
\begin{subfigure}{0.32\textwidth}
\includegraphics[width=\textwidth]{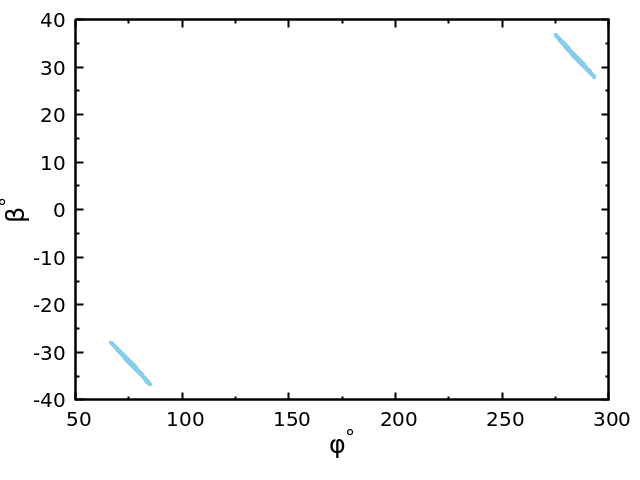}
\caption{}
\label{fig:5f}
\end{subfigure}
\captionsetup{justification=raggedright,singlelinecheck=false}
\caption{ (a) Variation of $m_1$, $m_2$, $m_3$ as a function of $\phi$, (b) variation of $\sum m_i$ as a function of $\phi$, (c) correlation 
between $\sum m_i$ and $M_{ee}$, (d) correlation between $\sum m_i$ and $m_{\nu}$,  (e) variation of $\alpha$ as a function of $\phi$, and (f)
variation of $\beta$ as a function of $\phi$ for TM$_1$ mixing matrix for Class $A_2$.}
\label{fig:5}
\end{figure}

\begin{figure}[htbp]
\begin{subfigure}{0.32\textwidth}
\includegraphics[width=\textwidth]{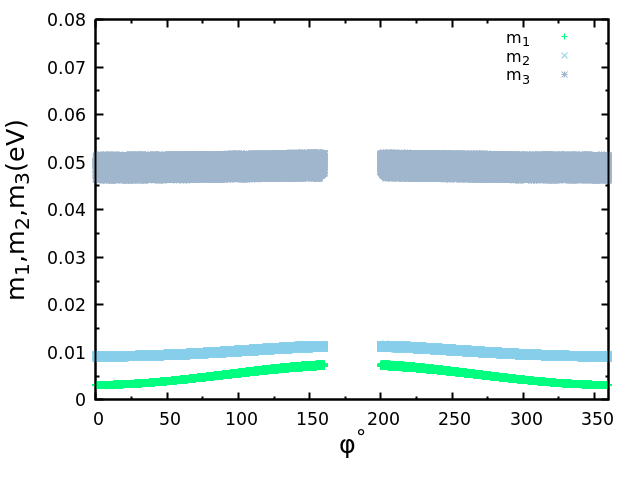}
\caption{}
\label{fig:6a}
\end{subfigure}
\hfill
\begin{subfigure}{0.32\textwidth}
\includegraphics[width=\textwidth]{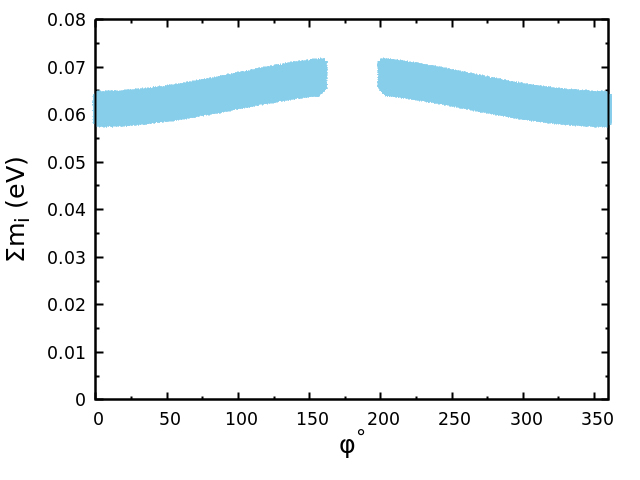}
\caption{}
\label{fig:6b}
\end{subfigure}
\hfill
\begin{subfigure}{0.32\textwidth}
\includegraphics[width=\textwidth]{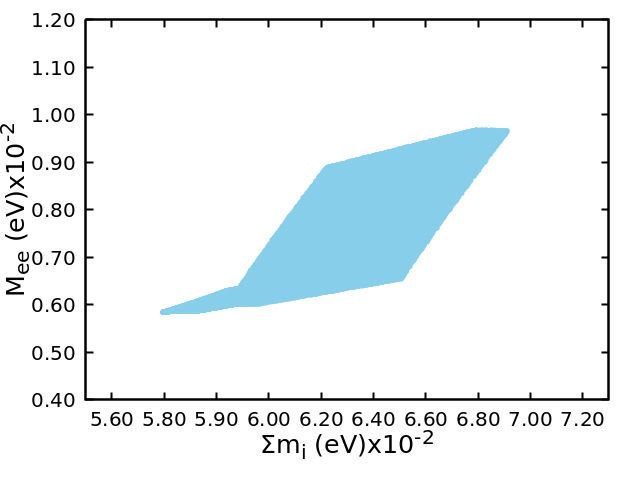}
\caption{}
\label{fig:6c}
\end{subfigure}
\hfill
\begin{subfigure}{0.32\textwidth}
\includegraphics[width=\textwidth]{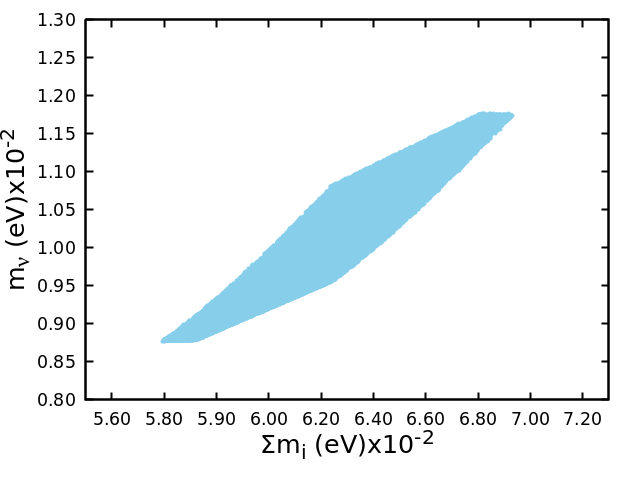}
\caption{}
\label{fig:6d}
\end{subfigure}
\hfill
\begin{subfigure}{0.32\textwidth}
\includegraphics[width=\textwidth]{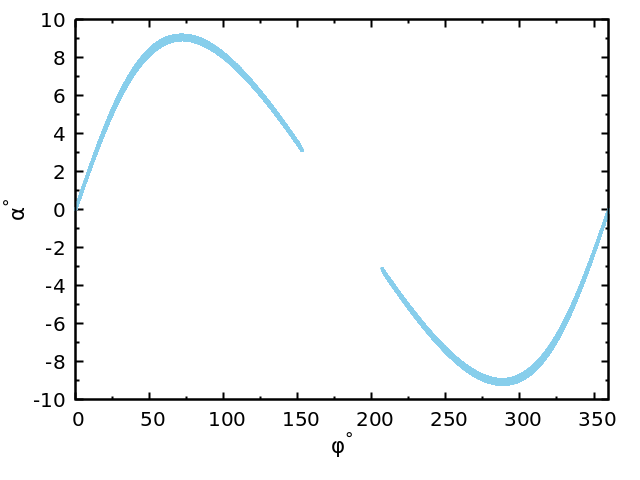}
\caption{}
\label{fig:6e}
\end{subfigure}
\hfill
\begin{subfigure}{0.32\textwidth}
\includegraphics[width=\textwidth]{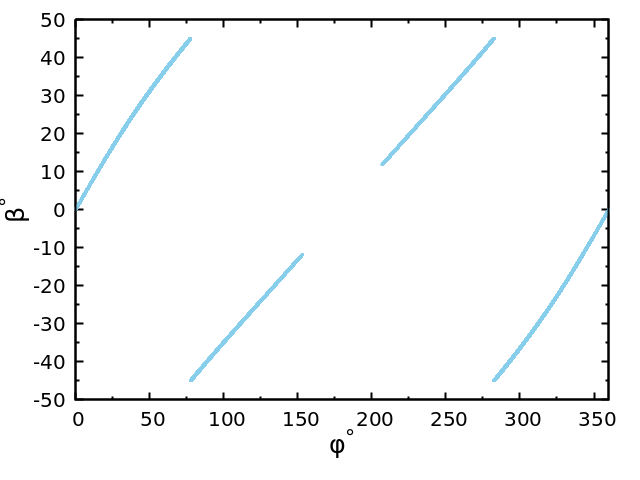}
\caption{}
\label{fig:6f}
\end{subfigure}
\captionsetup{justification=raggedright,singlelinecheck=false}
\caption{ (a) Variation of $m_1$, $m_2$, $m_3$ as a function of $\phi$, (b) variation of $\sum m_i$ as a function of $\phi$, (c) correlation 
between $\sum m_i$ and $M_{ee}$, (d) correlation between $\sum m_i$ and $m_{\nu}$,  (e) variation of $\alpha$ as a function of $\phi$, and (f)
variation of $\beta$ as a function of $\phi$ for TM$_2$ mixing matrix for Class $A_2$.}
\label{fig:6}
\end{figure}

Let us now proceed to discuss the phenomenological implication of class $A_2$ pattern on neutrino masses, the effective Majorana mass, 
effective electron anti-neutrino mass and the CP violating Majorana phases. We show the variation of neutrino masses $m_1$, $m_2$ and $m_3$ as 
a function of $\phi$ in Fig~\ref{fig:5a} and Fig.~\ref{fig:6a} for the TM$_1$ and the TM$_2$ mixing matrix, respectively.
In each case, they show normal mass ordering. The variation of $\sum m_i$ as a function of $\phi$ is shown in Fig~\ref{fig:5b} and 
Fig.~\ref{fig:6b} for TM$_1$ and TM$_2$ mixing matrix, respectively. In Fig.~\ref{fig:5c} and Fig.~\ref{fig:6c}, we have shown the 
correlation of $M_{ee}$ and $\sum m_i$ for both the mixing matrix, respectively. We have shown the correlation of $m_{\nu}$ with $\sum m_i$ 
for TM$_1$ and TM$_2$ mixing matrix in Fig.~\ref{fig:5d} and Fig.~\ref{fig:6d}, respectively. Also in Fig.~\ref{fig:5e} and Fig.~\ref{fig:6e},
we have shown the variation of $\alpha$ with respect to $\phi$ for both TM$_1$ and TM$_2$ mixing matrix, respectively. In Fig.~\ref{fig:5f} 
and Fig.~\ref{fig:6f}, we have shown the variation of $\beta$ with respect to $\phi$ for both the mixing matrix. The phenomenology of this 
class is quite similar to class A$_{1}$.

The best fit values and the corresponding allowed ranges of the absolute neutrino mass scale, the effective Majorana mass , the effective 
electron anti-neutrino mass and Majorana CP violating phases for this class are listed in Table.~\ref{tab:6}. The best fit values and
the allowed ranges of each observables obtained for this class are quite similar to the values obtained for class $A_1$.
\begin{table}[htbp!]
\begin{tabular}{|c|c|c|c|c|c|c|}
 \hline
 Mixing&Values& $\sum m_i\,{(\rm eV)}$ &$M_{ee}\,{(\rm eV)}$&$m_{\nu}\,{(\rm eV)}$&$\alpha^\circ$&$\beta^\circ$\\
 matrix&& &&&&\\
\hline
\multirow{2}{*}{TM$_1$}&Best fit & $0.066$ & $0.006$&$0.010$ &$\pm 9.58$& $\pm 30.73$\\
\cline{2-7}
&$3\sigma$ Range & $[0.060, 0.073]$ & $[0.006, 0.007]$&$[0.009, 0.012]$ &$[(-10.03,-8.84),\,(8.84, 10.03)]$& $[(-36.79,-27.82),\,(27.82, 36.79)]$\\
\hline
\multirow{2}{*}{TM$_2$}&Best fit & $0.064$ & $ 0.006$&$0.009$ &$\pm 7.84$&$\pm 29.21$\\
\cline{2-7}
&$3\sigma$ Range & $[0.057, 0.071]$ & $[0.006, 0.009]$&$[0.008, 0.012]$ &$[0,\pm 9.18]$&$[0,\pm 44.99]$\\
\hline
\hline
\end{tabular}
\caption{Best fit and $3\sigma$ allowed range of $\sum m_{i}(\rm eV)$, $M_{ee}(\rm eV)$, $m_{\nu}(\rm eV)$, $\alpha^\circ$ and $\beta^\circ$ 
for Class $A_2$.}
\label{tab:6}
\end{table}

\subsection{\textbf{Other Classes}}
The classes $B_3$, $B_4$, $B_5$, $B_6$, $S_1$, $S_2$, $F_1$, $F_2$ and $F_3$ are not acceptable for both TM$_1$ and TM$_2$ mixing matrix 
because the value of $r$ obtained for these classes are not within the experimental range. Similarly, the classes $S_3$, $F_4$ and $F_5$ are 
not allowed since they predict $m_1=m_2$ for both TM$_1$ and TM$_{2}$ mixing matrix, respectively. Moreover, the class $D$ is also not 
allowed as this class predicts $m_3=m_2$ for TM$_1$ mixing matrix and $m_1=m_3$ for TM$_2$ mixing matrix, respectively.

\section{Symmetry realization}
\label{section:5}
We can realize the symmetry of two minor zero in neutrino mass matrix through type-I seesaw  mechanism along with Abelian symmetry. For 
generating leptonic mass matrix, we need three right handed neutrinos $\nu_{Rp}$ $(p=1, 2, 3)$, three right handed charged leptons $l_{Rp}$ 
and three left handed lepton doublets $D_{Lp}$. Along with these, we need Higgs doublet for non zero elements $(M_l)_{pq}$ or $(M_D)_{pq}$, 
and scalar singlet for non zero elements $(M_R)_{pq}$ where $q = 1,2,3$. Higgs doublets get vacuum expectation values~(vevs) at the 
electroweak scale and the scalar singlets get the vevs at the seesaw 
scale. We follow the procedure discussed in Refs.~\cite{Lavoura:2004tu,Lashin:2011dn} to present the symmetry realization of class $A_1$ 
using $Z_{8}$ symmetry group. Under $Z_{8}$ symmetry, the leptons of first family remain invariant, leptons of second family changes sign and 
leptons of third family get multiplied by $\omega$ = ${\rm exp}(\frac{i\pi}{4})$.

The leptonic fields under $Z_{8}$ transform as
\begin{eqnarray}
\label{eq:50} 
&&\bar l_{R1} \rightarrow \bar l_{R1}\,, \qquad\qquad
\bar \nu_{R1} \rightarrow \bar \nu_{R1}\,, \qquad\qquad
D_{L1} \rightarrow \bar D_{L1}\,, \nonumber \\
&&\bar l_{R2} \rightarrow \omega^4 \bar l_{R2}\,, \qquad\qquad 
\bar \nu_{R2} \rightarrow \omega^4 \bar \nu_{R2}\,, \qquad\qquad  
D_{L2} \rightarrow \omega^4 \bar D_{L2}\,, \nonumber \\ 
&&\bar l_{R3} \rightarrow \omega \bar l_{R3}\,, \qquad\qquad 
\bar \nu_{R3} \rightarrow \omega \bar \nu_{R3}\,, \qquad\qquad  
D_{L3} \rightarrow \omega^7 \bar D_{L3}\,,
\end{eqnarray}
The bilinears $\bar l_{Rp}\,D_{Lq}$ and $\bar \nu_{Rp}\,D_{Lq}$ relevant 
for $(M_l)_{pq}$ and $(M_D)_{pq}$ transform as
\begin{equation}
\label{eq:51}
\bar l_{Rp}\,D_{Lq} \cong \bar \nu_{Rp}\,D_{Lq} \cong {
 \begin{pmatrix}
  1& \omega^4 & \omega^7\\
 \omega^4 & 1 &\omega^3\\
 \omega & \omega^5 & 1
\end{pmatrix}}\,
\end{equation}
and the bilinears $\bar \nu_{Rp} \bar \nu_{Rq}$ relevant for $(M_R)_{pq}$ transform as 
\begin{equation}
\label{eq:52}
\bar \nu_{Rp} \bar \nu_{Rq} \cong {
 \begin{pmatrix}
  1& \omega^4 & \omega\\
 \omega^4 & 1&\omega^5\\
 \omega & \omega^5 & \omega^2
\end{pmatrix}}\,.
\end{equation}

Under these transformation the diagonal Dirac mass matrices are generated automatically for both charged leptons and neutrinos. Also, under 
$Z_8$ transformation the elements $(1,1)$ and $(2,2)$ of Majorana mass matrix $M_R$ remain invariant. We can obtain non zero elements 
$(1,2)$ by introducing real scalar field $\chi_{12}$ which changes sign under $Z_8$ transformation and $(1,3)$ by introducing complex scalar 
fields $\chi_{13}$ which under $Z_8$ transformation gets multiplied by $\omega^7$ for $M_R$. In the absence of any scalar singlets other 
elements of $M_R$ remain zero .
The Majorana mass matrix $M_R$ can be written as
\begin{equation}
\label{eq:53}
 M_{R}= \begin{pmatrix}
1& b & c\\
 b & 1 &0\\
 c & 0 & 0
\end{pmatrix}\,.
\end{equation}
This gives two minor zero conditions corresponding to Class $A_1$ in the neutrino mass matrix. Class $A_2$ can also be realised similarly 
for different $M_R$.

\section{Conclusion}
\label{section:6}
We explore the consequences of two minor zeros in the neutrino mass matrix using trimaximal mixing matrix. There are total fifteen possible
patterns and out of these only two patterns namely class $A_1$ and class $A_2$ are found to be phenomenologically acceptable for TM$_1$ and 
TM$_2$ mixing matrix, respectively. We perform a naive $\chi^2$ analysis to find the best fit values of the two unknown parameters
$\theta$ and $\phi$ of the TM mixing matrix. We include five observables in our $\chi^2$ analysis namely, the three mixing 
angles~$(\theta_{13}, \theta_{12}, \theta_{23})$ and the two mass squared differences~$(\Delta m_{21}^2, \Delta m_{32}^2)$. It is found that
class $A_2$ with TM$_1$ mixing matrix provides the best fit to the experimental results. We also discuss the degree of fine tuning in the elements of the mass matrix for both classes $A_1$ and $A_2$  by introducing a new parameter $d_{FT}$. The $d_{FT}$ value is quite large in case of TM$_{1}$ mixing matrix for both the classes although the 
$\chi^2_{min}$ value obtained for Class  $A_2$ is quite small. Based on the $d_{FT}$ value, one can conclude that the degree of fine tuning
in the mass matrix element is quite strong in case of TM$_{1}$ mixing matrix.
The two allowed classes show normal mass ordering. We also give prediction of several unknown parameters such as the absolute neutrino mass scale, 
the effective Majorana mass, the effective electron anti-neutrino mass, three CP violating phases and the Jarlskog invariant measure of CP
violation. The effective Majorana mass obtained for each pattern is within the reach of neutrinoless
double beta decay experiment. The upper bound on the effective electron anti-neutrino mass obtained for each pattern is beyond the reach of 
current $\beta$ decay experiments.  
Moreover, we also discuss the symmetry realization of Class $A_1$ in the framework of type-I seesaw model 
using Abelian symmetry group $Z_8$.

\appendix
\section{}
\label{app}
The elements of neutrino mass matrix using TM$_1$ mixing matrix can be written as  
\begin{eqnarray} 
\label{eq:55}
  &&(M)_{ee}=\frac{2}{3}\,m_1+\frac{1}{3}\cos^2\theta\,m_2\,e^{2i\alpha}+\frac{1}{3}\sin^2\theta\,m_3\,e^{2\,i\,\beta},  
\nonumber \\ 
  &&(M)_{ e\mu}=(-\frac{1}{3})m_1+(\frac{1}{3}\cos^2\theta-\frac{1}{\sqrt{6}}\sin\theta \cos\theta e^{i\phi})m_2 e^{2i\alpha}+
(\frac{1}{3}\sin^2\theta+\frac{1}{\sqrt{6}}\sin\theta \cos\theta e^{i\phi}) m_3 e^{2i\beta},  \nonumber \\
  &&(M)_{ e\tau}=(-\frac{1}{3})m_1+(\frac{1}{3}\cos^2\theta+\frac{1}{\sqrt{6}}\sin\theta \cos\theta e^{i\phi})m_2 e^{2i\alpha}+(\frac{1}{3}\sin^2\theta-\frac{1}{\sqrt{6}}\sin\theta \cos\theta e^{i\phi}) m_3 e^{2i\beta},  \nonumber \\
  &&(M)_{ \mu\mu}=\frac{1}{6}m_1+(\frac{1}{\sqrt{3}}\cos\theta-\frac{1}{\sqrt{2}}\sin\theta  e^{i\phi})^2m_2 e^{2i\alpha}+
(\frac{1}{\sqrt{3}}\sin\theta+\frac{1}{\sqrt{2}}\cos\theta e^{i\phi})^2 m_3 e^{2i\beta},  \nonumber \\
  &&(M)_{ \mu\tau}=\frac{1}{6}m_1+(\frac{1}{3}\cos^2\theta-\frac{1}{2}\sin^2\theta e^{2i\phi}) m_2 e^{2i\alpha}+(\frac{1}{3}
\sin^2\theta-\frac{1}{2}\cos^2\theta e^{2i\phi}) m_3 e^{2i\beta},  \nonumber \\
  &&(M)_{ \tau\tau}=\frac{1}{6} m_1+(\frac{1}{\sqrt{3}}\cos\theta+\frac{1}{\sqrt{2}}\sin\theta e^{i\phi})^2 e^{2i\alpha}+
(\frac{1}{\sqrt{3}}\sin\theta-\frac{1}{\sqrt{2}}\cos\theta e^{i\phi})^2 m_3 e^{2i\beta}.  
 \end{eqnarray}

Using TM$_2$ mixing matrix, we can write the elements of neutrino mass matrix as   
   \begin{eqnarray} 
\label{eq:56}
 &&(M)_{ee}=(\frac{2}{3}\cos^2\theta)\,m_1+\frac{1}{3}\,m_2\,e^{2i\alpha}+(\frac{2}{3}\sin^2\theta)\,m_3\,e^{2\,i\,\beta},  
\nonumber \\ 
  &&(M)_{ e\mu}=(-\frac{1}{3}\cos^2\theta+\frac{1}{\sqrt{3}}\sin\theta \cos\theta e^{-i\phi}) m_1+\frac{1}{3}m_2 e^{2i\alpha}+
(-\frac{1}{3}\sin^2\theta-\frac{1}{\sqrt{3}}\sin\theta \cos\theta e^{-i\phi}) m_3 e^{2i\beta},  \nonumber \\
  &&(M)_{ e\tau}=(-\frac{1}{3}\cos^2\theta-\frac{1}{\sqrt{3}}\sin\theta \cos\theta e^{-i\phi}) m_1+\frac{1}{3}m_2 e^{2i\alpha}+
(-\frac{1}{3}\sin^2\theta+\frac{1}{\sqrt{3}}\sin\theta \cos\theta e^{-i\phi}) m_3 e^{2i\beta},  \nonumber \\
  &&(M)_{ \mu\mu}=(-\frac{1}{\sqrt{6}}\cos\theta+\frac{1}{\sqrt{2}}\sin\theta  e^{-i\phi})^2 m_1+\frac{1}{3}m_2 e^{2i\alpha}+
(\frac{1}{\sqrt{6}}\sin\theta+\frac{1}{\sqrt{2}}\cos\theta e^{-i\phi})^2 m_3 e^{2i\beta},  \nonumber \\
  &&(M)_{ \mu\tau}=(\frac{1}{6}\cos^2\theta-\frac{1}{2}\sin^2\theta e^{-2i\phi}) m_1+\frac{1}{3}m_2 e^{2i\alpha}+(\frac{1}{6}
\sin^2\theta-\frac{1}{2}\cos^2\theta e^{-2i\phi}) m_3 e^{2i\beta},  \nonumber \\
  &&(M)_{ \tau\tau}=(\frac{1}{\sqrt{6}}\cos\theta+\frac{1}{\sqrt{2}}\sin\theta  e^{-i\phi})^2 m_1+\frac{1}{3}m_2 e^{2i\alpha}+
(-\frac{1}{\sqrt{6}}\sin\theta+\frac{1}{\sqrt{2}}\cos\theta e^{-i\phi})^2 m_3 e^{2i\beta}.  
 \end{eqnarray}

\end{document}